\documentclass[12pt, letterpaper]{article}
\usepackage[pdftex]{graphicx}
\usepackage{epsfig}
\usepackage{latexsym}
\usepackage{amssymb}
\usepackage{amsmath}
\usepackage{wrapfig}
\usepackage{boxedminipage}
\usepackage{setspace}
\usepackage{subfigure}

\def\N{\cal N}

\DeclareSymbolFont{AMSb}{U}{msb}{m}{n}
\DeclareMathSymbol{\IN}{\mathbin}{AMSb}{"4E}
\DeclareMathSymbol{\IZ}{\mathbin}{AMSb}{"5A}
\DeclareMathSymbol{\IR}{\mathbin}{AMSb}{"52}
\DeclareMathSymbol{\Q}{\mathbin}{AMSb}{"51}
\DeclareMathSymbol{\II}{\mathbin}{AMSb}{"49}
\DeclareMathSymbol{\IC}{\mathbin}{AMSb}{"43}
\DeclareMathSymbol{\IP}{\mathbin}{AMSb}{"50}
\DeclareMathSymbol{\IH}{\mathbin}{AMSb}{"48}
\DeclareMathSymbol\IA{\mathalpha}{AMSb}{"41}
\DeclareMathSymbol\IS{\mathalpha}{AMSb}{"53}

\def\Q{{\cal Q}}

\def\N{{\cal N}}

\begin{document}
\begin{flushright}
\phantom{{\tt arXiv:0802.????}}
\end{flushright}


\begin{center} 
{\Large \bf  Universal Holographic Chiral Dynamics\\ in an External Magnetic Field}
  \end{center}

\bigskip \bigskip \bigskip

\centerline{\bf Veselin G. Filev${}^{\star}$, Clifford V. Johnson${}^{\dagger}$, Jonathan P. Shock${}^{\ddagger}$}

\bigskip
\bigskip
\bigskip

\centerline{${}^{\star}${\it School of Theoretical Physics, Dublin Institute For Advanced Studies}}
\centerline{\it 10 Burlington Road, Dublin 4, Ireland}
\centerline{\small \tt  filev@stp.dias.ie}
\bigskip
 
  \centerline{${}^{\dagger}$\it Department of Physics and Astronomy, University of Southern California}
\centerline{\it Los Angeles, CA 90089-0484, U.S.A.}
\centerline{\small \tt  johnson1@usc.edu}
\bigskip

\centerline{${}^{\ddagger}$\it Departamento de F\'\i sica de Part\'\i culas},
\centerline{\it Universidade de Santiago de Compostela and Instituto Galego de F\'\i sica de Altas Enerx\'\i as (IGFAE)}
\centerline{\it E-15782, Santiago de Compostela, Spain }
\centerline{\small \tt  shock@fpaxp1.usc.es}

\bigskip

\bigskip
\bigskip


\begin{abstract} 
\noindent 
In this work we further extend the investigation of holographic gauge
theories in external magnetic fields, continuing earlier work. We
study the phenomenon of magnetic catalysis of mass generation in 1+3
and 1+2 dimensions, using D3/D7-- and D3/D5--brane systems,
respectively. We obtain the low energy effective actions of the
corresponding pseudo Goldstone bosons and study their dispersion
relations. The D3/D7 system exhibits the usual
Gell-Mann--Oakes--Renner (GMOR) relation and a relativistic dispersion
relation, while the D3/D5 system exhibits a quadratic non-relativistic
dispersion relation and a modified linear GMOR relation. The low
energy effective action of the D3/D5 system is related to that
describing magnon excitations in a ferromagnet. We also study
properties of general Dp/Dq systems in an external magnetic field and
verify the universality of the magnetic catalysis of dynamical
symmetry breaking.

\end{abstract}
\newpage \baselineskip=18pt \setcounter{footnote}{0}
\section{Introduction}
The applications of holographic gauge/gravity correspondences to the
study of more and more diverse phenomena are ever widening in
scope. Over the last half a decade the links between finite
temperature generalizations of AdS/CFT and experimental heavy-ion
collisions have become much more concrete and the theoretical methods
available to us are yielding ever deeper results concerning the
properties of the quark-gluon plasma (see ref.~\cite{Edelstein:2009,Gubser:2009md}
for a recent review). Moreover in the last year the links between
holography and condensed matter systems have also flourished, with
work on superconductivity, and superfluidity, quantum phase
transitions and both the classical and quantum hall effects having
recent successes (e.g. refs.~\cite{KeskiVakkuri:2008eb,Hartnoll:2009sz}
and references therein).

In the present work we extend the investigation of holographic gauge
theories in the presence of external magnetic fields from the work
first studied in ref.~\cite{Filev:2007gb}. In this paper we are interested
in finding both universal properties of strongly coupled gauge
theories in the presence of magnetic fields, as well as in the
different phenomena exhibited in such theories in a variety of
space-time dimensions.

The phenomenon of dynamical flavor symmetry breaking catalysed by an
arbitrarily weak magnetic field is known from
refs.~\cite{Gusynin:1994re, Gusynin:1994xp} and refs.~\cite{Klimenko:1990rh, Klimenko:1991he, Klimenko:1992ch}. This effect was shown to
be model independent and therefore insensitive to the microscopic
physics underlying the low energy effective theory. In particular the
infra-red (IR) description of the Goldstone modes associated with the
dynamically broken symmetry should be universal. We therefore expect
to be able to study this phenomenon using the holographic
formalism. The aim of the present study will be to investigate the
dynamics of the Goldstone modes and construct the low energy chiral
Lagrangian of theories both in 3+1 and 2+1 dimensions in the presence
of external magnetic fields, showing that the appropriate holographic
models give precisely the results expected from the traditional field
theory approach.
 
The effective dynamics of fermion pairing, in $d+1$ dimensions, in the
presence of an external magnetic field is constrained to $d-2$ spatial
dimensions. For this reason there are marked differences in the
phenomenology of such systems in two and three spatial dimensions. In
2+1 dimensions refs.~\cite{Gusynin:1994xp}-\cite{Klimenko:1992ch} Poincare symmetry is broken
by the magnetic field (there is no longer any trace of the original
boost invariance), removing the strong constraints on the dynamics of
Goldstone modes imposed by special relativity. The naive Goldstone
boson counting therefore does not hold and the resulting dispersion
relation for the Goldstone modes takes a quadratic form, unlike in the
case of 3+1 dimensions where an $SO(1,1)$ subgroup of $SO(3,1)$
constrains the dynamics. Although the number of Goldstone particles is
no longer constrained in the non-relativistic setting, the number of
Goldstone fields is fixed by the dimension of ${G}/{H}$ ($G$=symmetry
of the action, $H$=symmetry of the ground state). We will show that
this also holds in the AdS/CFT context.

The 2+1 dimensional model is of particular interest because, as shown
in ref.~\cite{Hofmann:1998pp}, the low energy effective description is
that of magnon excitations in a ferromagnet. Using a D3/D5 brane
intersection we will be able to reproduce such a result at quadratic
order in the chiral Lagrangian. Moreover such 2+1 dimensional theories
may have relevance in the arenas of the quantum hall effect, and high
$T_c$ superconductivity.

In addition to the phenomena discussed specifically in two and three
spatial dimensions we show that certain universal behaviors are
exhibited holographically in the present context. Here we will study
holographic systems T-dual to the D3/D7 flavor model and show that the
existence of an arbitrarily small magnetic field induces a spiral
behaviour in the equation of state for such systems. In the limit that
the chiral symmetry of the underlying theory is preserved, this
equation of state can be studied analytically and such a symmetric
vacuum can be shown to be unstable. This is holographically equivalent
to the findings of refs.~\cite{Gusynin:1994re}-\cite{Klimenko:1992ch} -- flavor symmetry
breaking is induced dynamically by the presence of a magnetic field.

The outline of the present paper is as follows:

In section 2 we will return to the D3/D7 brane intersection in the
presence of an external magnetic field, discussed in
ref.~\cite{Filev:2007gb}. We shall show explicitly how the magnetic
catalysis of flavor symmetry breaking is realised in the holographic
system, including the calculation of the chiral Lagrangian to second
order in the low energy degrees of freedom. We will show that the
Gell-Mann--Oakes--Renner relation holds analytically and obtain the
dispersion relation for the Goldstone modes.

In section 3 we turn to the case of the D3/D5 defect theory and show
here how the $SO(3)$ flavor symmetry is dynamically broken to the
$U(1)$ subgroup in the presence of a magnetic field. In this
non-relativistic system we find the Goldstone modes and show that the
number of massless modes is not the same as the number of broken
generators, but satisfies a more general counting
rule~\cite{Nielsen:1975hm} applicable for non-relativistic systems. We
also show that the single Goldstone mode satisfies a modified
Gell-Mann--Oakes--Renner relation and a quadratic dispersion
relation. Again we obtain the dispersion relation analytically in the
small mass limit and find the low energy effective Lagrangian which
describes magnon excitations in a ferromagnet.

In section 4 we prove that the magnetic catalysis of dynamical
symmetry breaking is a universal effect in gauge theories dual to
Dp/Dq intersections in the appropriate decoupling limit. This proof
involves showing that all such systems exhibit a self--similar spiral
behaviour in their equation of state which leads to an instability for
the solution with zero dynamical mass. Just as in the work of
ref.~\cite{Gusynin:1994xp} this effect is independent of the magnitude
of the external magnetic field.
\section{Mass generation in the D3/D7 system}
In this section we will review the results of
refs.~\cite{Filev:2007gb,Filev:2007qu,Erdmenger:2007bn}, where a
holographic study of flavored ${\cal N}=4$ supersymmetric Yang-Mills
in an external magnetic field was studied using the D3/D7 system. We
will focus on the effect of mass generation by magnetic catalysis in
this theory and provide a detailed analysis of the pseudo-Goldstone
mode associated to the spontaneous breaking of a global $U(1)$
R-symmetry.  In particular we will show that the Gell-Mann--Oakes--Renner
relation for the mass of the corresponding $\eta'$ meson is satisfied.

The D3/D7 system provides a dual holographic description of $N_f$
fundamental $\N=2$ hypermultiplets coupled to $\N=4$ $SU(N_c)$
supersymmetric Yang Mills theory in the quenched approximation
$N_f\ll N_c$~\cite{Karch:2002sh}.  At zero separation between the D3
and D7--branes the fundamental hypermultiplets are massless and the
$\beta$--function of the theory is proportional to $N_f/N_c$. Thus in the
quenched approximation the $\beta$--function vanishes and the
corresponding gauge theory is conformal. The global $SO(6)$ R-symmetry
of the $\N=4$ SYM theory is broken to an $SU(2)\times U(1)$
R--symmetry, the $U(1)$ corresponding to rotations in the 2-plane
transverse to both the D3 and D7--branes. The left and right handed
fermions of the hypermultiplet have opposite charges under this
$U(1)_R$ and thus the formation of a fermionic condensate
$\langle\bar\psi\psi\rangle$ would lead to the spontaneous breaking of
this symmetry.

\subsection{Spontaneous symmetry breaking}

There are various ways in which one can study the breaking of the
chiral symmetry holographically. This has been studied in the past by
the deformation of AdS$_5\times S^5$ by a field corresponding to a
marginally irrelevant operator on the gauge theory side refs.~\cite{Babington:2003vm, Kruczenski:2003uq, Evans:2004ia}. In the
present case however we will stimulate the formation of a condensate
by turning on the magnetic components of the $U(1)$ gauge field of the
D7--branes $F_{\alpha\beta}$ (equivalent to exciting a pure gauge
$B-$field in the supergravity background). This $U(1)$ gauge field
corresponds to the diagonal $U(1)$ of the full $U(N_f)$ gauge symmetry
of the stack of D7--branes. Since the D7--branes wrap an infinite
internal volume, the dynamics of the $U(N_f)$ gauge field is frozen in
the four dimensional theory and the $U(N_f)$ gauge symmetry becomes a
global flavor symmetry $U(N_f)=U(1)_B\times SU(N_f)$. Therefore the
$U(1)$ gauge field that we consider corresponds to the gauged $U(1)_B$
baryon symmetry and the magnetic field that we introduce couples to
the baryon charge of the fundamental fields \cite{Myers:2007we}.

The problem thus boils down to studying embeddings of probe D7--branes
in the AdS$_5\times S^5$ background parameterized as follows:
\begin{eqnarray}
ds^2&=&\frac{\rho^2+L^2}{R^2}[ - dx_0^2 + dx_1^2 +dx_2^2 + dx_3^2 ]+\frac{R^2}{\rho^2+L^2}[d\rho^2+\rho^2d\Omega_{3}^2+dL^2+L^2d\phi^2]\ ,\nonumber\\
d\Omega_{3}^2&=&d\psi^2+\cos^2\psi d\beta^2+\sin^2\psi d\gamma^2, \label{geometry1}\\
g_{s}C_{(4)}&=&\frac{u^4}{R^4}dx^0\wedge dx^1\wedge dx^2 \wedge dx^3;~~~e^\Phi=g_s;~~~R^4=4\pi g_{s}N_{c}\alpha'^2\ ,\nonumber
\end{eqnarray}
where $\rho, \psi, \beta,\gamma$ and $L,\phi$ are polar coordinates in
the transverse $\mathbb{R}^4$ and $\mathbb{R}^2$ planes respectively.

Here $x_{a=1..3},\rho,\psi,\beta,\gamma$ parameterize the world volume
of the D7--brane and the following ansatz is considered for its
embedding:
\begin{eqnarray}
\phi\equiv {\rm const}\ ,\quad L\equiv L(\rho)\nonumber \label{anzatsEmb}\ ,
\end{eqnarray}
leading to the following induced metric on its worldvolume:
\begin{equation}
d\tilde s=\frac{\rho^2+L(\rho)^2}{R^2}[ - dx_0^2 + dx_1^2 +dx_2^2
+dx_3^2]+\frac{R^2}{\rho^2+L(\rho)^2}[(1+L'(\rho)^2)d\rho^2+\rho^2d\Omega_{3}^2] \  .
\label{inducedMetric}
\end{equation}
The D7--brane probe is described by the DBI action:
\begin{eqnarray}
S_{\rm{DBI}}=-N_f\mu_{7}\int\limits_{{\cal M}_{8}}d^{8}\xi e^{-\Phi}[-{\rm det}(G_{ab}+B_{ab}+2\pi\alpha' F_{ab})]^{1/2}\  . \label{DBI}
\end{eqnarray}

Here $\mu_{7}=[(2\pi)^7\alpha'^4]^{-1}$ is the D7--brane tension,
$G_{ab}$ and $B_{ab}$ are the induced metric and $B$-field on the
D7--brane's world volume, while $F_{ab}$ is its world--volume gauge
field. A simple way to introduce a magnetic field is to consider
a pure gauge $B$--field along the $x_2,x_3$ directions:
\begin{equation}
B^{(2)}= Hdx_{2}\wedge dx_{3} \label{anzats}\ .
\end{equation}
Since $B_{ab}$ and $F_{ab}$ appear on equal footing in the DBI action,
the introduction of such a $B$-field is equivalent to introducing an
external magnetic field of magnitude $H/(2\pi\alpha')$ to the dual
gauge theory.

Though the full solution of the embedding can only be calculated
numerically, the large $\rho$ behaviour (equivalently the ultraviolet
(UV) regime in the gauge theory language) can be extracted
analytically:
\begin{equation}
L(\rho)=m+\frac{c}{\rho^2}+\cdots \ .
\end{equation}
As discussed in ref.~\cite{Kruczenski:2003uq}, the parameters $m$ (the
asymptotic separation of the D7- and D3- branes) and $c$ (the degree
of bending of the D7--brane in the large $\rho$ region) are related to
the bare quark mass $m_{q}=m/2\pi\alpha'$ and the fermionic condensate
$\langle\bar\psi\psi\rangle\propto -c$ respectively. It should be
noted that the boundary behavior of $L(r)$ really plays the role of
source and vacuum expectation value (vev) for the full ${\cal N}=2$
hypermultiplet of operators. In the present case, where supersymmetry
is broken by the gauge field configuration, we are only interested in
the fermionic bilinears and this will refer only to quarks, and not
their supersymmetric counterparts.

\begin{figure}[h] 
   \centering
   \includegraphics[width=10cm]{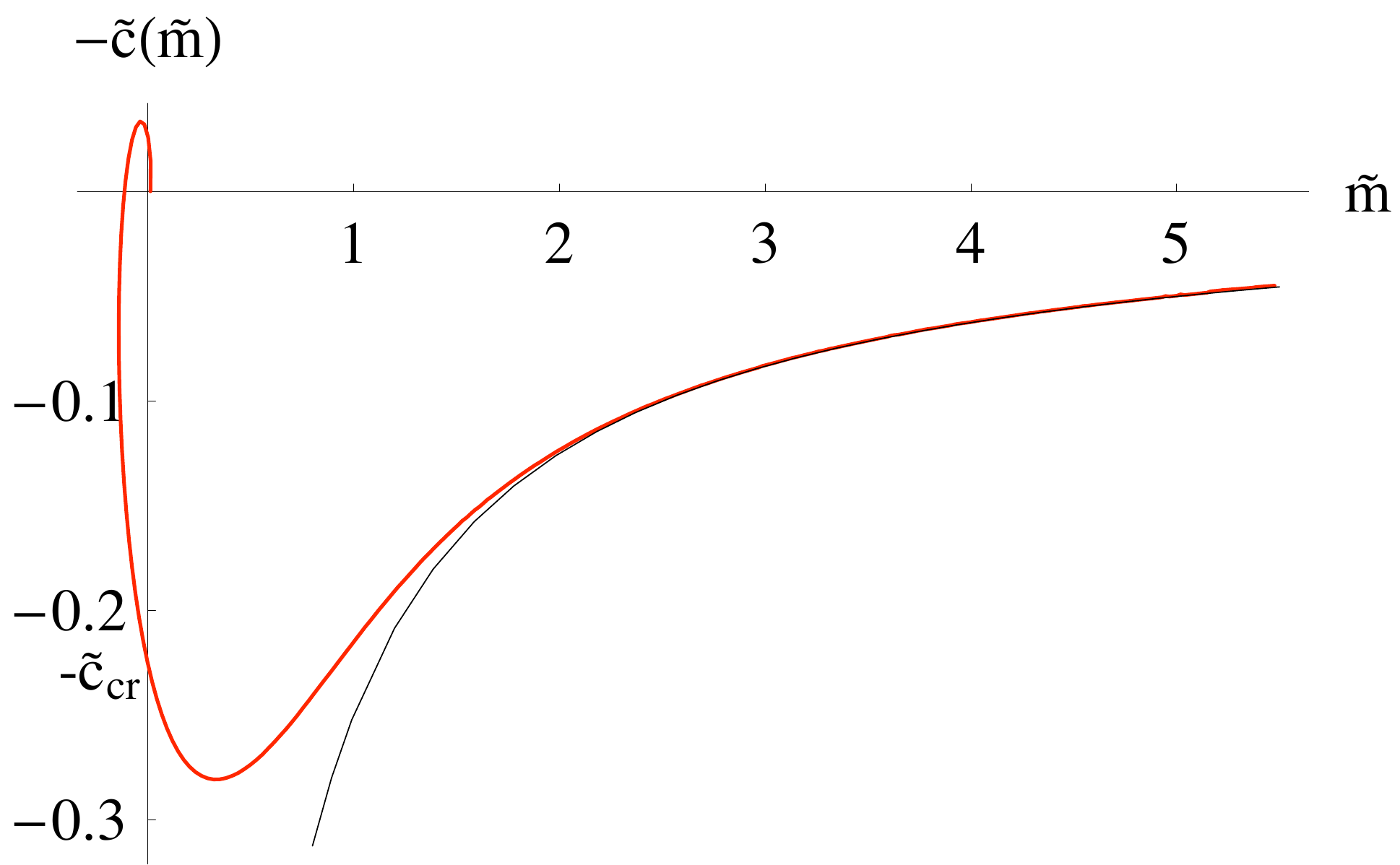}
   \caption{\small Parametric plot of $\tilde{c}$ against $\tilde{m}$
     for fundamental matter in the presence of an external magnetic
     field. The lower (black) line represents the curve $1/\tilde m$,
     fitting the large $\tilde{m}$ behavior. It is also evident that
     for the outer branch of the spiral, for $\tilde m=0$ the
     condensate, $\langle\bar\psi\psi\rangle$ is non-zero. The
     corresponding value of the condensate is $\tilde c_{\rm
       cr}=0.226$.}  \label{fig:fig1}
\end{figure}

At this point it is convenient to introduce dimensionless parameters
$\tilde c=c/R^3H^{3/2}$ and $\tilde m=m/R\sqrt{H}$. By performing a
numerical shooting method from the infrared while varying the small
$\rho$ boundary value, $L(\rho\rightarrow 0)=L_{IR}$, we recover the
parametric plot presented in figure~\ref{fig:fig1}, the main result
explored in ref.~\cite{Filev:2007gb}.

The lower (black) curve corresponds to the analytic behavior of
$\tilde c(\tilde m)=1/\tilde m$ for large $\tilde m$. The most
important observation is that at $\tilde m=0$ there is a non-zero
fermionic condensate:
\begin{equation}
\langle\bar\psi\psi\rangle=-\frac{N_fN_c}{(2\pi\alpha')^3\lambda}c=-\frac{N_fN_c\tilde c_{\rm{cr}}}{(2\pi^2)^{3/4}\lambda^{1/4}}\left(\frac{H}{2\pi\alpha'}\right)^{3/2}\ .\label{cond}
\end{equation}

Where $\lambda=g_{YM}^2N_c$ is the 't Hooft coupling and $\tilde
c_{\rm{cr}}\approx 0.226$ is a numerical constant corresponding to the
$y$-intercept of the outer spiral from figure~\ref{fig:fig1}.
Equation (\ref{cond}) is telling us that the theory has developed a
negative condensate that scales as
$\left(\frac{H}{2\pi\alpha'}\right)^{3/2}$. This is not surprising,
since the theory is conformal in the absence of the scale introduced
by the external magnetic field.  The energy scale controlled by the
magnetic field, $\left(\frac{H}{2\pi\alpha'}\right)^{1/2}$, leads to an
energy density proportional to
$\left(\frac{H}{2\pi\alpha'}\right)^{2}$. In order to lower the
energy, the theory responds to the magnetic field by developing a
negative fermionic condensate.
 
\begin{figure}[h] 
   \centering
   \includegraphics[width=12cm]{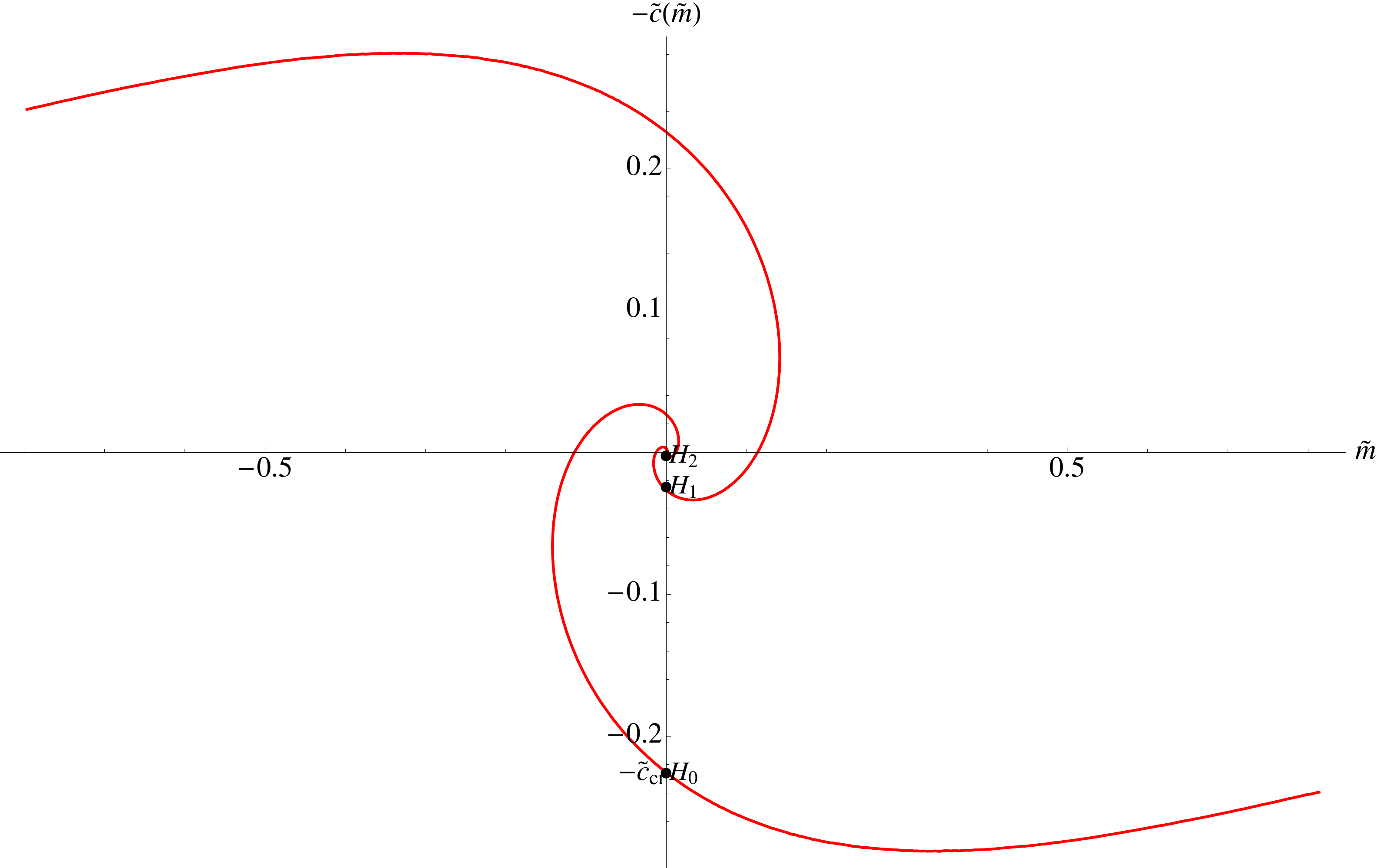}
   \caption{\small A magnification of figure~\ref{fig:fig1} shows the spiral behavior near the origin of the $(-\tilde c,\tilde m)$-plane. The second (left) spiral arm represents the $(\tilde m, -\tilde c)\to (-\tilde m,\tilde c)$ symmetry of the theory.}
   \label{fig:fig2}
\end{figure}

Another interesting feature of the theory is the discrete--self--similar
structure of the equation of state ($\tilde c$ {\it vs.} $\tilde m$) in the
vicinity of the trivial $\tilde m=0$ embedding, namely the origin of
the plot from figure~\ref{fig:fig1} presented in figure~\ref{fig:fig2}.

This double logarithmic structure has been analyzed in
ref.~\cite{Filev:2007qu}, where a study of the meson spectrum revealed
that only the outer branch of the spiral is tachyon free and
corresponds to a stable phase having spontaneously broken chiral
symmetry. In Section 3 of this paper we will show that an identical
structure is also present for the D3/D5 system and in Section 4 we
will demonstrate that this structure is a universal feature of the
magnetic catalysis of mass generation for gauge theories
holographically dual to Dp/Dq intersections.

A further result of
refs.~\cite{Filev:2007gb,Filev:2007qu,Erdmenger:2007bn} was the
detailed analysis of the light meson spectrum of the theory. In
ref.~\cite{Filev:2007gb} it was shown that the introduction of an
external magnetic field breaks the degeneracy of the spectrum studied
in ref.~\cite{Kruczenski:2003be}. This manifests itself as Zeeman
splitting of the energy levels. In the limit of zero quark mass, the
study also revealed the existence of a massless ``$\eta'$ meson"
corresponding to the spontaneously broken $U(1)_R$ symmetry. In the
next subsection we will revisit the study of the meson spectrum of the
theory and provide an analytic proof of the Gell-Mann--Oakes--Renner
relation \cite{GellMann:1968rz}:
 \begin{equation} 
M_{\pi}^2=-\frac{2\langle\bar\psi\psi\rangle}{f_{\pi}^2}m_q\ ,\label{GellMann1}
 \end{equation} 
 in the spirit of the analysis performed in
 ref.~\cite{Kruczenski:2003uq}.
\subsection{The Gell-Mann--Oakes--Renner relation - an analytic derivation}

In order to study the light meson spectrum of the theory one needs to
consider the quadratic fluctuations of the D7--brane embedding and
study the corresponding normal modes
\cite{Kruczenski:2003be}. Technically one should consider the full
supergravity action for the D7--branes:
\begin{equation}
S_{\rm{tot}}=S_{\rm{DBI}}+S_{\rm{WZ}}\ ,\label{ful}
\end{equation}
where $S_{\rm{DBI}}$ is given by equation (\ref{DBI}) and the relevant part of the Wess-Zumino term is given by \cite{Filev:2007gb}:
\begin{equation}
S_{\rm{WZ}}=\frac{(2\pi\alpha')^2}{2}\mu_{7}\int{F_{(2)}\wedge F_{(2)}\wedge C_{(4)}}+(2\pi\alpha')\mu_{7}\int F_{(2)}\wedge B_{(2)}\wedge \tilde P[C_{(4)}]\ ,
\label{WZ}
\end{equation}
The next step is to consider fluctuations of the D7--brane in the transverse $\mathbb{R}^2$:
\begin{eqnarray}
L=L_0(\rho)+2\pi\alpha'\delta L\ ;~~~\phi=2\pi\alpha'\Phi\ ,
\end{eqnarray}
and expand equation~(\ref{ful}) to second order in $\alpha'$. Note
that with such an expansion we should also consider fluctuations of
the $U(1)$ gauge field on the D7--brane. As demonstrated in
refs.~\cite{Filev:2007gb,Arean:2005ar} the effect of the magnetic
field will be to mix the equations of motion for the scalar and vector
fluctuations. In particular $\Phi$ couples to the $A_0$ and $A_1$
components of the gauge field, while $\delta L$ couples to the $A_2$
and $A_3$ components. The rest of the components of the vector field
decouple and can be consistently set to zero. This splitting of the
meson spectrum is a manifestation of the broken Lorentz
symmetry. Indeed the external magnetic field breaks the $SO(1,3)$
Lorentz symmetry down to $SO(1,1)\times SO(2)$ corresponding to boosts
in the $x_0,x_1$ plane and rotations in the $x_2,x_3$ plane. Since the
massless ``pion" that we are interested in corresponds to fluctuations
along $\phi$, we will excite only the $\Phi,A_0,A_1$ fields. The
relevant terms of the expansion are \cite{Filev:2007gb}:

\begin{eqnarray}
&&{\cal L}_{\phi\phi}=-(2\pi\alpha')^2\frac{\mu_7}{g_s}\frac{1}{2}\sqrt{|g_{S^3}|}\frac{gR^2L_0^2}{\rho^2+L_0^2}S^{ab}\partial_a\Phi\partial_b\Phi\ , \label{qvd2}\\
&&{\cal L}_{\Phi A}=-(2\pi\alpha')^2\frac{\mu_7}{g_s}\sqrt{|g_{S^3}|}H\partial_{\rho}K\Phi F_{01}\ ,\nonumber\\
&&{\cal L}_{AA}=-(2\pi\alpha')^2\frac{\mu_7}{g_s}\sqrt{|g_{S^3}|}\frac{1}{4}gS^{aa'}S^{bb'}F_{ab}F_{a'b'}\ ,\nonumber
\end{eqnarray}

where:

\begin{eqnarray}
&&||S^{ab}||={\rm diag}\{-G_{11}^{-1},G_{11}^{-1},\frac{G_{11}}{G_{11}^{2}+H^2},\frac{G_{11}}{G_{11}^{2}+H^2},G_{\rho\rho}^{-1},G_{\psi\psi}^{-1},G_{\alpha\alpha}^{-1},G_{\beta\beta}^{-1}\}\ ,\label{S}\\
&&g(\rho)=\rho^3\sqrt{1+{L_0}'^2}\sqrt{1+\frac{R^4H^2}{(\rho^2+L_0^2)^2}};~~~K(\rho)=\frac{R^4\rho^4}{(\rho^2+L_0^2)^2};~~~\sqrt{|g_{S^3}|}=\sin\psi\cos\psi\ .\nonumber
\end{eqnarray}

Here $L_0(\rho)$ corresponds to the classical embedding of the D7--brane and $G_{ab}$ are the components of the background metric equation~(\ref{geometry1}).

The equations of motions for $\Phi$ and $F_{01}$ are calculated from the quadratic action, resulting in:

\begin{eqnarray}
&&\frac{1}{g(\rho)}\partial_{\rho}\left(\frac{g(\rho)L_0^2\partial_{\rho}\Phi}{1+L_0'^2}\right)+\frac{L_0^2\Delta_{\Omega_3}\Phi}{\rho^2}+\frac{R^4L_0^2}{(\rho^2+L_0^2)^2}{\widetilde\Box}\Phi-\frac{H\partial_{\rho}K}{g(\rho)}F_{01}=0\ ,\label{EOM1}\\
&&\frac{1}{g(\rho)}\partial_{\rho}\left(\frac{g(\rho)\partial_{\rho}F_{01}}{1+L_0'^2}\right)+\frac{\Delta_{\Omega_3}F_{01}}{\rho^2}+\frac{R^4}{(\rho^2+L_0^2)^2}{\widetilde\Box} F_{01}-\frac{H\partial_{\rho}K}{g(\rho)}(-\partial_0^2+\partial_1^2)\Phi=0\nonumber\ ,
\end{eqnarray}
where $F_{01}=\partial_0A_1-\partial_1A_0$ and the gauge constraint $-\partial_0A_0+\partial_1A_1=0$ is imposed (note that this is the usual Lorentz gauge, corresponding to the unbroken $SO(1,1)$) and we have defined:
\begin{equation}
\widetilde\Box=-\partial_0^2+\partial_1^2+\frac{\partial_2^2+\partial_3^2}{1+\frac{R^4H^2}{(\rho^2+L_0^2)^2}}\ .\label{Box}
\end{equation}

\begin{figure}[h] 
   \centering
   \includegraphics[width=9cm]{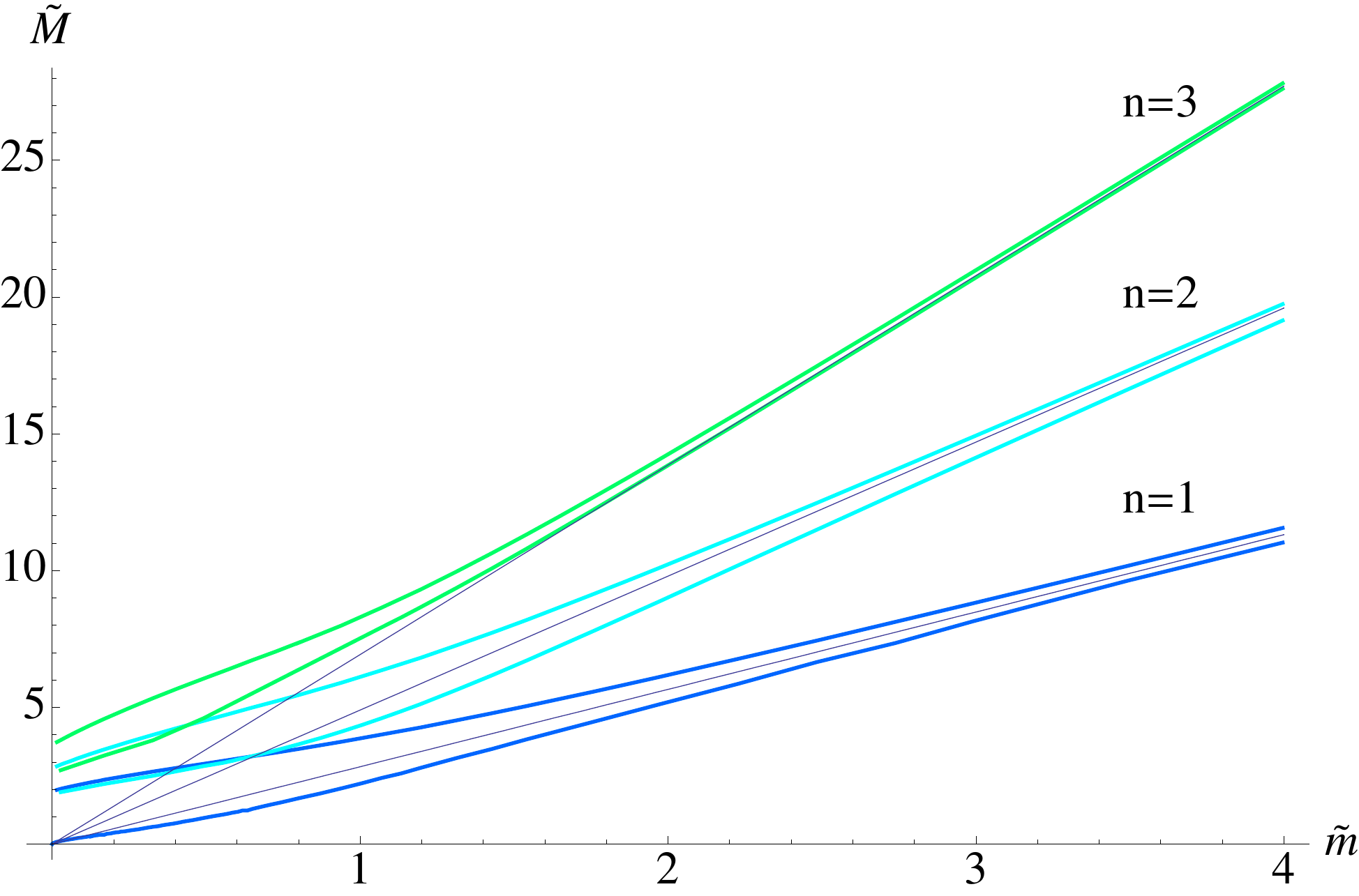}
   \caption{\small There is  Zeeman splitting of the states
     due to the magnetic field. In the absence of the field there are
     three straight lines emanating from the origin; these are split
     to form six curves. At zero bare quark
     mass (the end of the lowest curve) there is indeed a massless
     Goldstone mode. The straight lines correspond to the asymptotic
     AdS results.}
   \label{fig:mesonD7}
\end{figure}

\begin{figure}[h] 
   \centering
   \includegraphics[width=9cm]{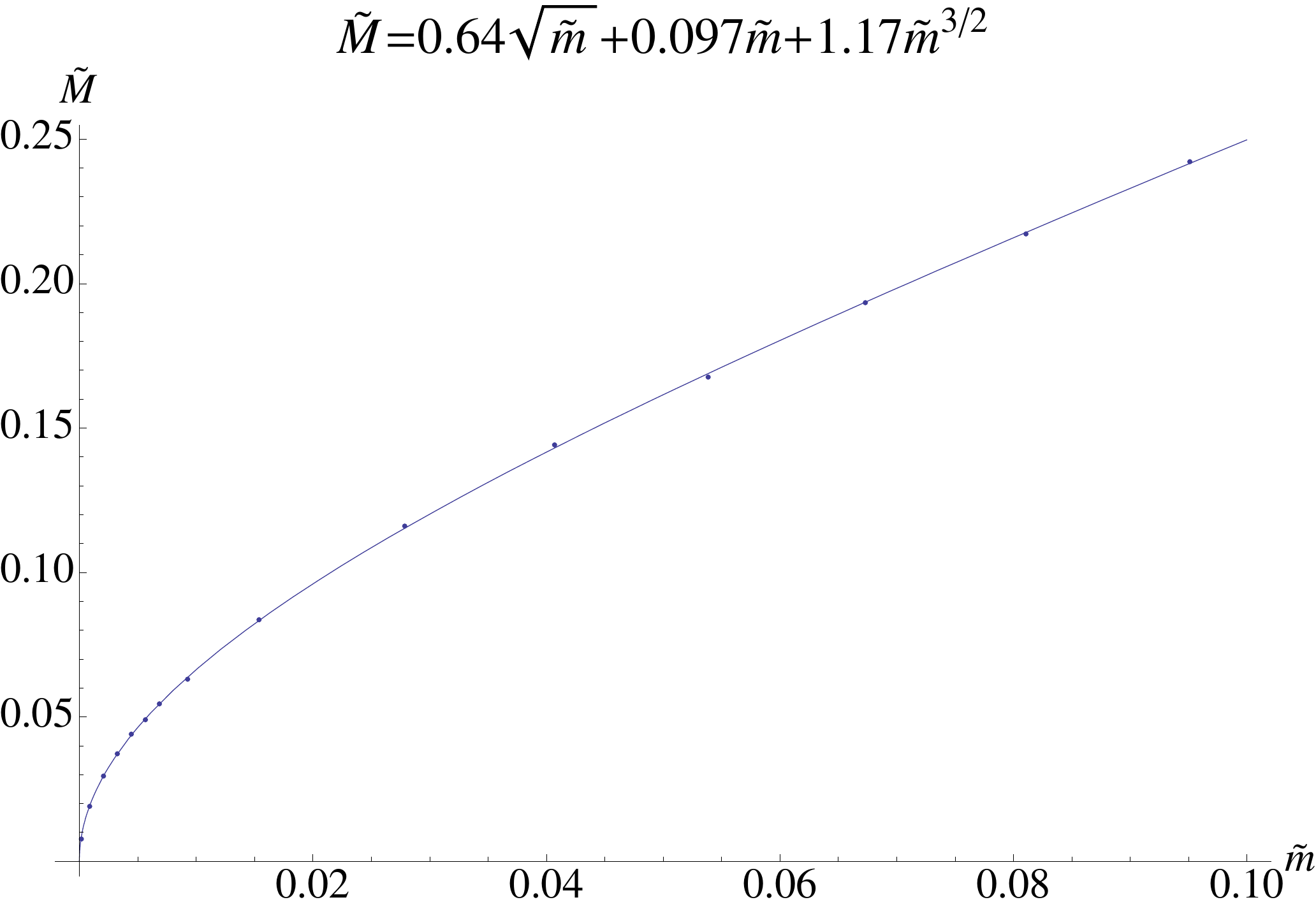}
   \caption{\small There is a characteristic $\tilde M\propto
     \sqrt{\tilde m}$ behavior at small bare quark mass.}
   \label{fig:zoomedD7}
\end{figure}

Once again the broken Lorentz symmetry is manifest in equation
(\ref{Box}). The definition of the spectrum is now a subtle issue in
the presence of the broken space-time symmetry. We will define the
spectrum as the energy of a particle as measured in its rest frame. In
fact because we retain the $SO(1,1)$ symmetry we may consider
fluctuations propagating in the $x_1$ direction. Since we are
interested in describing the lowest lying modes (``pions'' in
particular) we will focus on modes that have no $S^3$
dependence. Therefore we consider the ans\"atze:
\begin{equation}
\Phi=e^{i(k_0x^0+k_1x^1)}h(\rho);~~~F_{01}=e^{i(k_0x^0+k_1x^1)}f(\rho)\label{anz1}\, ,
\end{equation}
and define:
\begin{equation}
M^2=k_0^2-k_1^2\ .
\end{equation}
The equations (\ref{EOM1}) simplify to:
\begin{eqnarray}
&&\frac{1}{g}\partial_{\rho}\left(\frac{gL_0^2}{1+L'^2_0}\partial_{\rho}h\right)+\frac{R^4L_0^2}{(\rho^2+L_0^2)^2}M^2h-\frac{H\partial_{\rho}K}{g}f=0\ ,\label{EOMsmpl} \\
&&\frac{1}{g}\partial_{\rho}\left(\frac{g}{1+L'^2_0}\partial_{\rho}f\right)+\frac{R^4}{(\rho^2+L_0^2)^2}M^2f-\frac{M^2H\partial_{\rho}K}{g}h=0\ .\nonumber
\end{eqnarray}
Note that for large bare masses $m$ (and correspondingly large values
of $L$) the term proportional to the magnetic field is suppressed and
the meson spectrum should approximate to the result for the pure
AdS$_5\times S^5$ space-time case studied in
ref.~\cite{Kruczenski:2003be}, where the authors obtained the
following relation:
\begin{equation}
M_n=\frac{2m}{R^2}\sqrt{(n+1)(n+3)}\ ,\label{purespect}
\end{equation}
between the eigenvalue of the $n^{th}$ excited state $\omega_n$ and the bare mass $m$. If one imposes the boundary conditions:
\begin{equation}
h(\epsilon)=1;~~~h'(\epsilon)=0;~~~f(\epsilon)=1;~~~f'(\epsilon)=0\ , 
\end{equation}
the coupled system of differential equations can be solved
numerically.  Then by requiring the functions $h(\rho)$ and $f(\rho)$
to be regular at infinity one can quantize the spectrum of the
fluctuations. It is also convenient to define the following
dimensionless parameter $\tilde M=MR/\sqrt{H}$. The resulting plot for
the first three excited states is presented in figure
\ref{fig:mesonD7}. There is Zeeman splitting of the states due to the
magnetic field. (In the absence of the field there are three straight
lines emanating from the origin; these are split to form six curves.)
Also, at zero bare quark mass there is indeed a massless Goldstone
mode, appearing at the end of the lowest curve. Furthermore the plot
in figure~\ref{fig:zoomedD7} shows that for small bare quark mass one
can observe a characteristic $\tilde M\propto \sqrt{\tilde m}$
dependence. In the next section we shall provide an analytic proof of
that relation and obtain an integral expression for the numerical
coefficient $0.64$ presented above the plot in
figure~\ref{fig:zoomedD7}.

In the following section we shall demonstrate that for small bare
quark mass, $m_q=m/2\pi\alpha'$, the spectrum exhibits the
characteristic $M^2\propto{m}$ dependence. Once we have illustrated
that the functional dependence is correct we will show that the
constant of proportionality is also that expected from the GMOR
relation. Furthermore we shall generalize the ans\"atze (\ref{anz1})
to consider fluctuations depending on both the momentum along the
magnetic field $\vec k_{||}=(k_1,0,0)$ and the transverse momentum
$\vec k_{\perp}=(0,k_2,k_3)$:
\begin{equation}
\Phi=e^{i(\omega t+\vec k . \vec x)}h(\rho)\ ;~~~F_{01}=e^{i(\omega t+\vec k.\vec x)}f(\rho)\label{anz2}\ .
\end{equation}
 We shall also show that for small $\omega=k_0$ and $|{\vec k}|$ the following dispersion relation holds:
\begin{equation}
\omega(\vec k)^2=M^2+\vec k_{||}^2+\gamma\vec k_{\perp}^2\ ;~~~\omega=k_0\ ;~~~\vec k_{||}=(k_1,0,0)\ ;~~~\vec k_{\perp}=(0,k_2,k_3)\ ,\label{disp-rel}
\end{equation}
where $\gamma$ is a constant that we shall determine.

\subsubsection{The $M^2\propto m$ dependence}

Using an approach similar to the one employed in ref.~\cite{Kruczenski:2003uq} we define:
\begin{eqnarray}
&&\Psi^2=\frac{gL_0^2}{1+L'^2_0}\ ;~~~\nu=R^4\frac{1+L'^2_0}{(\rho^2+L_0^2)^2}\ ;~~~\tilde\nu=R^4\frac{1+L'^2_0}{(\rho^2+L_0^2)^2}\frac{1}{1+\frac{R^4H^2}{(\rho^2+L_0^2)^2}}\ ,\label{PSIS}\\ 
&&\Psi_1=\Psi/L_0\ ;~~~\psi=h\Psi\ ;~~~\psi_1=f\Psi_1\ .\nonumber
\end{eqnarray}
The equations of motions (\ref{EOM1}) can then be written in the compact form:
\begin{eqnarray}
&&\ddot\psi-\frac{\ddot\Psi}{\Psi}\psi=-(\omega^2-\vec k_{||}^2)\nu\psi+\vec k_{\perp}^2\tilde\nu\psi+\frac{H\partial_{\rho}K}{\Psi\Psi_1}\psi_1\ ,\label{EOMc}\\
&&\ddot\psi_1-\frac{\ddot\Psi_1}{\Psi_1}\psi_1=-(\omega^2-\vec k_{||}^2)\nu\psi_1+\vec k_{\perp}^2\tilde\nu\psi_1+\frac{H\partial_{\rho}K}{\Psi\Psi_1}(\omega^2-\vec k_{||}^2)\psi\nonumber \ .
\end{eqnarray}
Let us remind the reader that for large $\rho$, $L_0(\rho)$ has the behavior:
\begin{equation}
L_0\propto m+\frac{c}{\rho^2}+\cdots\ ,\label{exp2}
\end{equation}
Let us denote by $\bar L_0$ the classical embedding corresponding to
$(m=0,c=c_{\rm{cr}})$. It is relatively easy to verify that at
$m=0,\vec k_{\perp}=\vec 0$ and correspondingly $M^2=\omega^2-\vec
k_{||}^2=0$ the choice:
\begin{equation}
\psi=\bar\Psi\equiv\Psi|_{\bar L_0}\ ; ~~~\psi_1=0\ ,
\end{equation}
is a solution to the system (\ref{EOMc}).  Next we consider embeddings
corresponding to a small bare quark mass $\delta m$. This will
correspond to small nonzero values of $M^2$ and $\vec k_{\perp}^2$. It
is then natural to consider the following variations:
\begin{eqnarray}
&&\psi=\bar\Psi+\delta\psi \label{variation}\ ,\\
&&\psi_1=0+\delta\psi_1\nonumber\ ,
\end{eqnarray}
where $\delta\psi$ and $\delta\psi_1$ are of order $M^2$. Note that
$M$ corresponds to the mass of the ground state at $m_q=\delta
m/2\pi\alpha'$ and we are assuming that the variations of the wave
functions $\delta\psi$ and $\delta\psi_1$ are infinitesimal for
infinitesimal $m_q$. After expanding in equation~(\ref{EOMc}) we get
the following equations of motion:
\begin{eqnarray}
  &&\delta\ddot\psi-\frac{\ddot{\bar\Psi}}{\bar\Psi}\delta\psi-\delta\left(\frac{\ddot{\Psi}}{\Psi}\right)\bar\Psi=-(\omega^2-\vec k_{||}^2)\bar\nu\bar\Psi+\vec k_{\perp}^2\bar{\tilde v}\bar\Psi+\frac{H\partial_{\rho}K}{\bar\Psi_1\bar\Psi}\delta\psi_1\ ,\label{EOMVAR}\\
  &&\bar\Psi_1\delta\ddot{\psi_1}-\ddot{\bar{\Psi}}_1\delta\psi_1=H\partial_{\rho} K(\omega^2-k_{||}^2)\nonumber\ ,
\end{eqnarray}
where $\bar{\nu}=\nu|_{\bar L_0}$. The second equation in (\ref{EOMVAR}) can be integrated to give:
\begin{equation}
{\bar\Psi}_1\delta\dot\psi_1-\dot{\bar\Psi}_1\delta\psi_1=HK(\omega^2-k_{||}^2)+\rm{constant}\ .
\end{equation}
From the boundary conditions that $K|_{\rho=0}=0$ and
$\bar\Psi_1(0)=0,\dot{\bar\Psi}_1(0)=0$ we see that the constant of
integration is zero and arrive at:
\begin{equation}
\partial_{\rho}\left(\frac{\delta\psi_1}{\bar\Psi_1}\right)=\frac{HK(\omega^2-k_{||}^2)}{\bar\Psi_1^2}\ .\label{der}
\end{equation}
Next we multiply the first equation in (\ref{EOMVAR}) by $\bar\Psi$ and integrate along $\rho$ to obtain:
\begin{eqnarray}
&&(\omega^2-\vec k_{||}^2)\int\limits_0^{\infty}d\rho\bar\nu\bar\Psi^2-\vec k_{\perp}^2\int\limits_0^{\infty}d\rho\bar{\tilde\nu}\bar\Psi^2=-\int\limits_0^{\infty}(\bar\Psi\delta\ddot\psi-\ddot{\bar\Psi}\delta\psi)d\rho+\int\limits_0^{\infty}{\bar\Psi}^2\delta\left(\frac{\ddot\Psi}{\Psi}\right)d\rho+\label{integr}\\ 
&&+H\int\limits_0^{\infty}\frac{\partial_{\rho}K\delta\psi_1}{\bar\Psi_1}d\rho=-(\bar\Psi\delta\dot\psi-\dot{\bar\Psi}\delta\psi)\Big |_0^{\infty}+(\bar\Psi\delta\dot\Psi-\dot{\bar\Psi}\delta\Psi)\Big |_0^{\infty}-H\int\limits_0^{\infty}K\partial_\rho\left(\frac{\delta\psi_1}{{\bar\Psi}_1}\right)d\rho\nonumber\ ,
\end{eqnarray}
where the last term on the right-hand side of equation~(\ref{integr})
has been integrated by parts using the fact that $\delta\psi_1$ should
be regular at infinity. From the definition of $\bar\Psi$ it follows
that $\bar\Psi\propto \rho^{3/2}L_0(0)$ as $\rho\to 0$ and
$\bar\Psi\propto c/\rho^{1/2}$ as $\rho\to\infty$. This together with
the requirement that $\psi_1$ is regular at $\rho=0$ and vanishes at
infinity, suggests that the first term on the right-hand side of
equation~(\ref{integr}) vanishes. For the next term, we use the fact
that:
\begin{equation}
\delta\Psi=\rho^{3/2}\delta\left({\frac{1+\frac{H^2R^4}{(\rho^2+L^2_0)^2}  }{1+L'^2_0}  }\right)^{1/4}L_0+\rho^{3/2}\left(\frac{1+\frac{H^2R^4}{(\rho^2+L^2_0)^2}  }{1+L'^2_0} \right)^{1/4}\delta L_0\ ,
\end{equation}
and therefore obtain:
\begin{eqnarray}
&&\delta\Psi|_0=0;\quad \delta\dot\Psi |_0=0\ ,\\ \nonumber
&&\delta\Psi |_{\infty}\propto\rho^{3/2}\delta m\ ;\quad\delta\dot\Psi|_{\infty}\propto \frac{3}{2}\sqrt{\rho}\delta m\ .
\end{eqnarray}
The second term in equation~(\ref{integr}) then becomes:
\begin{equation}
(\bar\Psi\delta\dot\Psi-\dot{\bar\Psi}\delta\Psi)\Big |_0^{\infty}=2c\delta m\ .
\end{equation}
Finally using the equality in equation~(\ref{der}) we arrive at the result:
\begin{equation}
(\omega^2-\vec k_{||}^2)\int\limits_0^\infty d\rho\left\{\bar\nu{\bar\Psi}^2+\frac{H^2{\bar K}^2}{{{\bar\Psi}^2_1}}\right\}-\vec k_{\perp}^2\int\limits_0^{\infty}d\rho\bar{\tilde\nu}\bar\Psi^2=2c\delta m\ .
\end{equation}
Now we define:
\begin{equation}
\gamma=\left(\int\limits_0^{\infty}d\rho\bar{\tilde\nu}\bar\Psi^2\right)/\left(\int\limits_0^\infty d\rho\left\{\bar\nu{\bar\Psi}^2+\frac{H^2{\bar K}^2}{{{\bar\Psi}^2_1}}\right\}\right)\ ,
\end{equation}
and solve for $M^2$ from equation (\ref{disp-rel}) to obtain:
\begin{equation}
M^2\int\limits_0^\infty d\rho\left\{\bar\nu{\bar\Psi}^2+\frac{H^2{\bar K}^2}{{{\bar\Psi}^2_1}}\right\}=2c\delta m\ .\label{GMOR1}
\end{equation}
Equation (\ref{GMOR1}) suggests that the mass of the ``pion"
associated to the softly broken global $U(1)$ symmetry satisfies the
Gell-Mann--Oakes--Renner relation \cite{GellMann:1968rz}:
 \begin{equation} 
M_{\pi}^2=-\frac{2\langle\bar\psi\psi\rangle}{f_{\pi}^2}m_q\ .\label{GellMann2}
 \end{equation}
 In order to prove equation~(\ref{GellMann2}) we need to evaluate the
 effective coupling of the ``pion" $f_{\pi}^2$. Noting that $\delta
 m\propto m_q$ and $c\propto -\langle\bar\psi\psi\rangle$, we conclude
 that:
 \begin{equation}
f_{\pi}^2\propto\int\limits_0^{\infty}d\rho\left\{\bar\nu{\bar\Psi}^2+\frac{H^2{\bar K}^2}{{{\bar\Psi}^2_1}}\right\}\ .
\end{equation}
At this point is useful to verify the consistency of our analysis by
comparing the coefficient in equation (\ref{GMOR1}) to the numerically
determined coefficient $0.64$ from the plot in
figure~\ref{fig:zoomedD7}. Indeed from equation~(\ref{GMOR1}) we
obtain:
\begin{equation}
\tilde M/\sqrt{\tilde m}=\left[\frac{1}{2\tilde c_{\rm{cr}}}\int\limits_0^{\infty}d\tilde\rho\left\{\bar{\hat\nu}{\bar{\hat\Psi}}^2+\frac{{\bar {\hat K}}^2}{{{\bar{\hat\Psi}}^2_1}}\right\}\right]^{-1/2}\approx 0.655\ ,
\end{equation}
where we have defined the dimensionless quantities:
\begin{equation}
\hat\nu=H^2\nu;~~\hat\Psi^2=\Psi^2/R^5H^{5/2};~~\hat\Psi_1^2=\Psi_1^2/R^3H^{3/2};~~\hat K=K/R^4\ .
\end{equation}
There is excellent agreement with the fit from
figure~\ref{fig:zoomedD7}.

Next we will obtain an effective four dimensional action for
the ``pion" and from this derive an exact expression for $f_\pi^2$.

\subsubsection{Effective chiral action and $f_{\pi}^2$}

In this section we will reduce the eight dimensional world-volume
action for the quadratic fluctuations of the D7--brane to an effective
action for the massless ``pion" associated to the spontaneously broken
global $U(1)$ symmetry. Note that our effective action should be
describing a single ``pion" mode, while the 8D action given by
equation~(\ref{qvd2}) describes the dynamics of two independent
degrees of freedom, namely $\Phi$ and $F_{01}$ coupled by the 
magnetic B-field {\it via} the second equation in equation~(\ref{qvd2}). As
rigid rotations along $\phi$ correspond to chiral rotations, (the
asymptotic value of $\phi$ at infinity corresponds to the phase of the
condensate in the dual gauge theory) the spectrum of $\Phi$ at zero
quark mass contains the Goldstone mode that we are interested in. This
is why we first integrate out the gauge field components $A_0$ and
$A_1$ and then dimensionally reduce to four dimensions.

Furthermore as mentioned earlier, because of the magnetic field the
$SO(1,3)$ Lorentz symmetry is broken down to $SO(1,1)\times SO(2)$
symmetry. This is why in order to extract the value of $f_{\pi}^2$ we
consider excitations of $\Phi$ depending only on the $x_0,x_1$
directions and read off the coefficient in front of the kinetic term. The resulting on-shell effective action for $\Phi$ is:
\begin{equation}
S^{\rm{eff}}=-{\cal{N}}\int d^4x\left[-(\partial_0\Phi)^2+(\partial_1\Phi)^2\right]\ ,
\end{equation}
where ${\cal N}$ is given by:
\begin{equation}
{\cal N}=(2\pi\alpha')^2\frac{\mu_7}{g_s}N_f\pi^2\int\limits_0^{\infty}d\rho\left\{\bar\nu{\bar\Psi}^2+\frac{H^2{\bar K}^2}{{{\bar\Psi}^2_1}}\right\}\ .
\end{equation}
We refer the reader to Appendix A for a detailed derivation of the 4D
effective action $S^{\rm{eff}}$.

We have defined $\Phi$ {\it via} $\phi=(2\pi\alpha')\Phi$, where $\phi$ corresponds to rotations in the transverse $\IR^2$ plane and is the angle of chiral rotation in the dual gauge theory. The chiral Lagrangian is then given by: 
\begin{equation}
S^{\rm{eff}}=-(2\pi\alpha')^2\frac{f_{\pi}^2}{4}\int d^4x \partial_{\mu}\Phi\partial^{\mu}\Phi\ ;~~~\mu=0\,\, {\rm or} \,\, 1\, ,
\end{equation}  
and therefore:
\begin{equation}
f_{\pi}^2=N_f 4\pi^2\frac{\mu_7}{g_s}\int\limits_0^\infty d\rho\left(\bar\nu\bar\Psi^2+\frac{H^2K^2}{{\bar\Psi}_1^2}\right)\label{Fpi}\ .
\end{equation}

The D7--brane charge in equation~(\ref{Fpi}) is given by
$\mu_7=[(2\pi)^7\alpha'^4]^{-1}$ and the overall prefactor in equation
(\ref{Fpi}) can be written as $N_fN_c/2(2\pi\alpha')^4\lambda$. Now,
recalling the expressions for the fermionic condensate,
equation~(\ref{cond}), and the bare quark mass, $m_q=m/2\pi\alpha'$,
one can easily verify that equation (\ref{GMOR1}) is indeed the
Gell-Mann--Oakes--Renner relation:
 \begin{equation} 
M_{\pi}^2=-\frac{2\langle\bar\psi\psi\rangle}{f_{\pi}^2}m_q\ .\label{GellMann3}
 \end{equation}

 It turns out that for small momenta $\vec k_{||}, \vec k_{\perp}$ and
 small mass $M_{\pi}^2$ one can obtain the following more general
 effective 4D action (see appendix A for a detailed derivation):
\begin{equation}
S_{\rm{eff}}=-{\cal N}\int d^4x\left\{[-(\partial_0\tilde\Phi)^2+(\partial_1\tilde\Phi)^2]+\gamma[(\partial_2\tilde\Phi)^2+(\partial_3\tilde\Phi)^2]-\frac{2\langle\bar\psi\psi\rangle}{f_{\pi}^2}m_q\tilde\Phi^2\right\}+\cdots\ ,\label{effacttext}
\end{equation}
where $\gamma$ is defined in equation (\ref{appendconst}). As one can
see, the action (\ref{effacttext}) is the most general quadratic
action consistent with the $SO(1,1)\times SO(2)$ symmetry and suggests
that pseudo Goldstone bosons satisfy the dispersion relation
(\ref{disp-rel}).

\section{Mass generation in the D3/D5 system}

In this section we provide a holographic description of the magnetic
catalysis of chiral symmetry breaking in $1+3$ dimensional $SU(N_c)$
$\N=4$ supersymmetric Yang-Mills theory coupled to~$N_f$ $\N=2$
fundamental hypermultiplets confined to a $1+2$ dimensional
defect. Recently this theory received a great deal of attention and
emphasis has been made of the potential application of this brane
configuration in describing qualitative properties of $1+2$
dimensional condensed matter systems (see for example
refs.~\cite{{Evans:2008zs},{O'Bannon:2008bz},{Myers:2008me}}). In this
section we will study the effect of an external magnetic field on the
theory and demonstrate that the system develops a dynamically
generated mass and negative fermionic condensate leading to a
spontaneous breaking of a global $SO(3)$ symmetry down to a $U(1)$
symmetry. On the gravity side this symmetry corresponds to the
rotational symmetry in the transverse $\IR^3$.  Naively there should
be two massless Goldstone bosons corresponding to the generators of
the coset $SO(3)/U(1)$. As we will show the 1+2 dimensional nature of
the defect theory leads to a coupling of the transverse scalars
corresponding to the coset generators and as a result there is only a
single Goldstone mode. Furthermore the characteristic
$M_{\pi}\propto\sqrt{m}$ Gell-Mann--Oakes--Renner relation is modified to a
linear $M_{\pi}\propto m$ behavior. It turns out that these features
can be understood from a low energy effective theory point of
view. Indeed in $1+2$ dimensions the effect of the magnetic field is
to break the $SO(1,2)$ Lorentz symmetry down to $SO(2)$ rotational
symmetry and as a result the theory is non-relativistic. A single time
derivative chemical potential term is allowed (there is no boost
symmetry) and interestingly the supergravity action generates such a
term through the Wess-Zumino contribution of the D5--brane. It is this
term that is responsible for the modified counting rule of the number
of Goldstone bosons \cite{Nielsen:1975hm} and leads to a quadratic
dispersion relation as well as to the modified linear
Gell-Mann--Oakes--Renner relation. Another interesting feature of the
model is that to quadratic order the effective low energy action is
the same as the effective action describing spin waves in a
ferromagnet \cite{Hofmann:1998pp} in an external magnetic field. We
comment briefly on the possible applications of this similarity.

\subsection{Generalities}

Let us consider the AdS$_5\times S^5$ supergravity background
(\ref{geometry1}) and introduce the following parameterization:
\begin{eqnarray}
ds^2&=&\frac{u^2}{R^2}[ - dx_0^2 + dx_1^2 +dx_2^2 + dx_3^2 ]+\frac{R^2}{u^2}[dr^2+r^2d\Omega_{2}^2+dl^2+l^2d\tilde\Omega_2^2]\ ,\\
u^2&=&r^2+l^2\ ;~~d\Omega_2^2=d\alpha^2+\cos^2\alpha d\beta^2\ ;~~d\tilde\Omega_2^2=d\psi^2+\cos^2\psi d\phi^2\ .\nonumber
\end{eqnarray}
We have split the transverse $\IR^6$ to $\IR^3\times\IR^3$ and
introduced spherical coordinates $r,\Omega_2$ and $l,\tilde\Omega_2$
in the first and second $\IR^3$ planes respectively. Next we introduce
a stack of probe $N_f$ D5--branes extended along the $x_0,x_1,x_2$
directions, and filling the $\IR^3$ part of the geometry parameterized
by $r,\Omega_2$. As mentioned above on the gauge theory side this
corresponds to introducing $N_f$ fundamental $\N=2$ hypermultiplets
confined on a $1+2$ dimensional defect. The asymptotic separation of
the D3 and D5 --branes in the transverse $\IR^3$ space parameterized
by $l$ corresponds to the mass of the hypermultiplet. In the following
we will consider the following anzatz for a single D5--brane:
\begin{equation}
l=l(r)\ ;~~~\psi=0\ ;~~~\phi=0\ .\label{anz2}
\end{equation}
The asymptotic separation $m=l(\infty)$ is related to the bare mass of
the fundamental fields {\it via} $m_q=m/2\pi\alpha'$. If the D3 and D5
branes overlap, the fundamental fields in the gauge theory are
massless and the theory has a global $SO(3)\times SO(3)$
symmetry. Clearly a non-trivial profile of the D5--brane $l(r)$ in the
transverse $\IR^3$ would break the global symmetry down to
$SO(3)\times U(1)$, where $U(1)$ is the little group in the transverse
$\IR^3$. If the asymptotic position of the D5--brane vanishes ($m=0$)
this would correspond to a spontaneous symmetry breaking, the non-zero
separation $l(0)$ on the other hand would naturally be interpreted as
the dynamically generated mass of the theory.

Note that the D3/D5 intersection is T--dual to the D3/D7 intersection
from the previous section and thus the system is supersymmetric. The
D3 and D5 --branes are BPS objects and there is no attractive potential
for the D5--brane, hence the D5--brane has a trivial profile $l\equiv
const$. However a non-zero magnetic field will break the supersymmetry
and as we are going to demonstrate, the D5--brane will feel an
effective repulsive potential that will lead to dynamical mass
generation. In order to introduce a magnetic field perpendicular to
the plane of the defect, we consider a pure gauge $B$-field in the
$x_1,x_2$ plane given by:
\begin{equation}
B=Hdx_1\wedge dx_2\ .
\end{equation}
This is equivalent to turning on a non-zero value for the $0,1$
component of the gauge field on the D5--brane. The magnetic field
introduced into the dual gauge theory in this way has a magnitude
$H/2\pi\alpha'$. The D5--brane embedding is determined by the DBI
action:
\begin{eqnarray}
S_{\rm{DBI}}=-N_f\mu_{5}\int\limits_{{\cal M}_{6}}d^{6}\xi e^{-\Phi}[-{\rm det}(G_{ab}+B_{ab}+2\pi\alpha' F_{ab})]^{1/2}\  . \label{DBI2}
\end{eqnarray}
Where $G_{ab}$ and $B_{ab}$ are the pull-back of the metric and the $B$-field respectively and $F_{ab}$ is the gauge field on the D5--brane.

With the anzatz (\ref{anz2}) the Lagrangian is given by:
\begin{equation}
{\cal L}\propto r^2\sqrt{1+l'^2}\sqrt{1+\frac{R^4H^2}{(r^2+l^2)^2}}\ .\label{lagr2}
\end{equation}
From this it is trivial to solve the equation of motion for $l(r)$
numerically, imposing $l(0)=l_{in}$ and $l'(0)$ as initial
conditions. Clearly, at large $r$ the Lagrangian (\ref{lagr2})
asymptotes to that at zero magnetic field and hence we have the
asymptotic solution \cite{Mateos:2007vn}:
\begin{equation}
l(r)=m+\frac{c}{r}+\cdots\ ,\label{exp3}
\end{equation}
where $c\propto\langle\bar\psi\psi\rangle$ the condensate of the fundamental fields. 

\subsection{Spontaneous symmetry breaking}

\begin{figure}[h] 
   \centering
   \includegraphics[width=9cm]{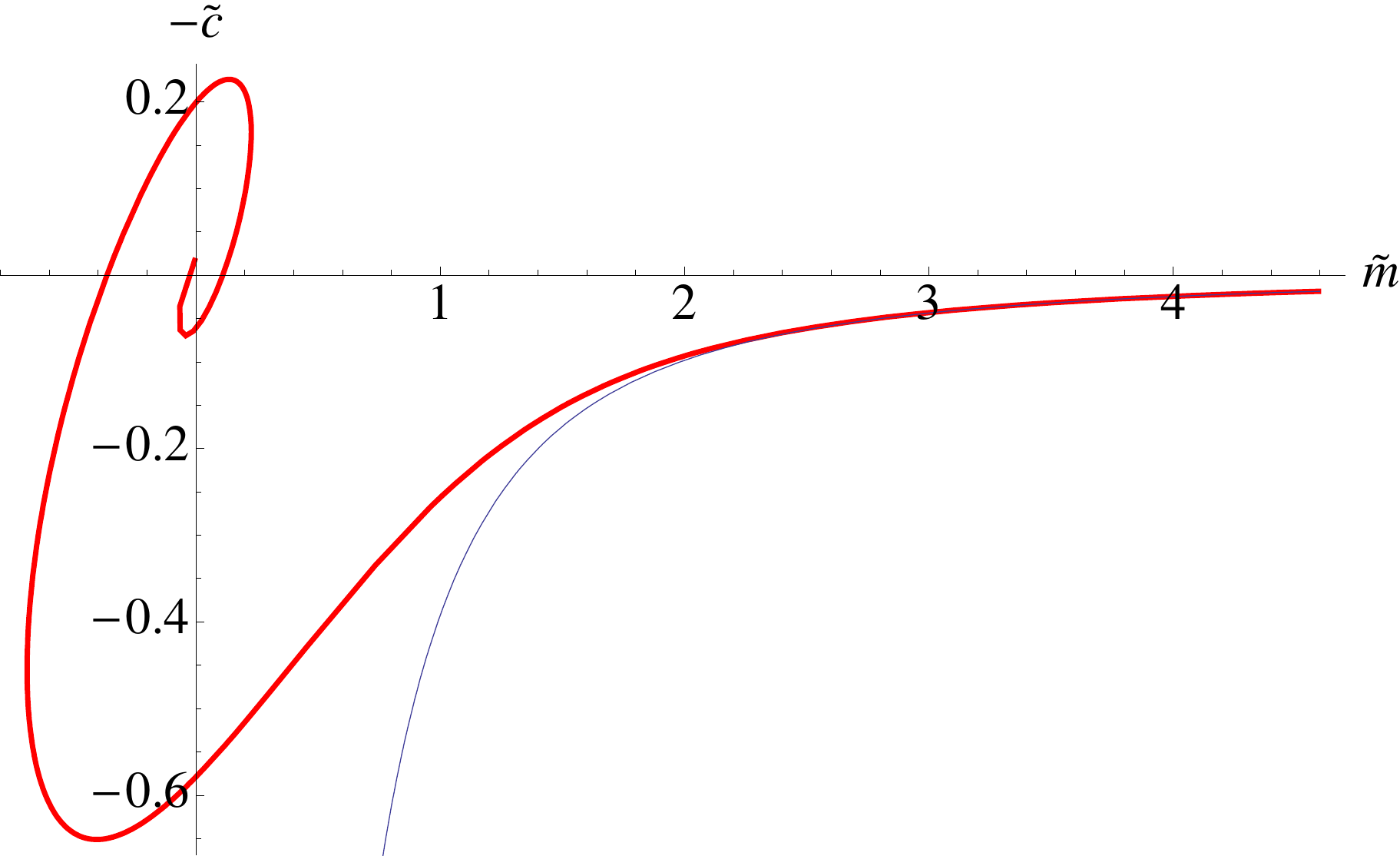}
   \caption{\small A plot of $-\tilde c$ {\it vs.} $\tilde m$. At zero
     bare mass $\tilde m=0$ the theory has developed a negative
     condensate $\langle\bar\psi\psi\rangle\propto -\tilde
     c_{\rm{cr}}\approx -0.59$. For large $\tilde m$ there is 
     excellent  agreement with equation~(\ref{largm}), as  represented by the
     lower (blue) curve. }
   \label{fig:fig3}
\end{figure}

Before solving the equation of motion it is convenient to introduce dimensionless variables:
\begin{equation}
\tilde r=r/R\sqrt{H}\ ;~~~\tilde l=l/R\sqrt{H}\ ;~~~\tilde m=m/R\sqrt{H}\ ;~~~\tilde c=c/R^2H\ .
\end{equation}
The Lagrangian (\ref{lagr2}) can then be written as:
\begin{equation}
{\cal L}\propto {\tilde r}^2\sqrt{1+{\tilde l}'^2}\sqrt{1+\frac{1}{(\tilde r^2+\tilde l^2)^2}}\ .
\end{equation}
The corresponding equation of motion is:
\begin{equation}
\partial_{\tilde r}\left(\frac{\tilde r^2l'}{\sqrt{1+\tilde l'^2}}\frac{\sqrt{1+(\tilde r^2+\tilde l^2)^2}}{
(\tilde r^2+\tilde l^2)}\right)=-2\frac{\tilde r^2\tilde l\sqrt{1+\tilde l'^2}}{(\tilde r^2+\tilde l^2)^2\sqrt{1+(\tilde r^2+\tilde l^2)^2}}\ .\label{EOM2}
\end{equation}
Before solving equation~(\ref{EOM2}) it will be useful to extract the
asymptotic behavior of $\tilde c(\tilde m)$ at large $\tilde m$. To
this end we use that at large $\tilde m$ the separation $\tilde
l(\tilde r)\approx \tilde m=\rm{const}$. The equation of motion then
simplifies to:
\begin{equation}
\partial_{\tilde r}(\tilde r^2\tilde l')=-\frac{2\tilde r^2\tilde m}{(\tilde r^2+\tilde m^2)^3}\ ,
\end{equation}
and hence:
\begin{equation}
\tilde r^2\tilde l'=-2\tilde m\int\limits_0^{\tilde r}d\tilde r\frac{\tilde r^2}{(\tilde r^2+\tilde m^2 )^2}\ .
\end{equation}
Using the expansion (\ref{exp3}) one can verify that:
\begin{equation}
\lim_{\tilde r \to +\infty}\tilde r^2\tilde l'=\tilde c=2\tilde m\int\limits_0^{\infty}d\tilde r\frac{\tilde r^2}{(\tilde r^2+\tilde m^2)^3}=\frac{\pi}{8\tilde m^2}\label{largm}\ .
\end{equation}
Equation (\ref{largm}) can thus be used as a check of the accuracy of
our numerical results. Indeed the numerically generated plot of
$-\tilde c$ {\it vs.} $\tilde m$ is presented in figure
\ref{fig:fig3}. The most important observation is that at zero bare
mass $\tilde m$ the theory has developed a negative condensate
$\langle\bar\psi\psi\rangle\propto -\tilde c_{\rm{cr}}\approx
-0.59$. It can also be seen that for large $\tilde m$ the numerically
generated plot is in good agreement with equation (\ref{largm})
represented by the lower (black) curve. Another interesting feature of the
equation of state is the spiral structure near the origin of the
parameter space analogous to the one presented in figure
\ref{fig:fig2} for the case of the D3/D7 system. We will come back to
this in Section 4 in more general terms, and show that this feature is
universal for the class of gauge theories dual to the Dp/Dq systems.

In order to show that the global $SO(3)$ symmetry is indeed
spontaneously broken we need to study the free energy of the
theory. Indeed the existence of the spiral structure suggests that
there is more than one phase at zero bare mass, corresponding to the
different $y$-intercepts of the $-\tilde c$ {\it vs.} $\tilde m$
plot. We will demonstrate below that the lowest positive branch of the
curve presented in figure~\ref{fig:fig3} is the stable one.

Following ref.~\cite{Mateos:2007vn} we will identify the regularized
wick rotated on-shell action of the D5--brane with the free energy of
the theory. Let us introduce a cut-off at infinity, $r_{\max}$, The
wick rotated on-shell action is given by:
\begin{equation}
S=N_f\frac{\mu_5}{g_s}4\pi V_3R^3H^{3/2}\int\limits_0^{\tilde r_{\rm{max}}} d\tilde r\tilde r^2\sqrt{1+\tilde l'^2}\sqrt{1+\frac{1}{(\tilde r^2+\tilde l^2)^2}}\ ,\label{actionwick}
\end{equation}
where $V_3=\int d^3x$ and $\tilde l(\tilde r)$ is the solution of
equation (\ref{EOM2}). It is easy to verify, using the expansion from
equation (\ref{exp3}),  that the integral in equation~(\ref{actionwick})
has the following behavior at large $\tilde r_{max}$:
\begin{equation} 
\int\limits_0^{\tilde r_{\rm{max}}} d\tilde r\tilde r^2\sqrt{1+\tilde l'^2}\sqrt{1+\frac{1}{(\tilde r^2+\tilde l^2)^2}}=\frac{1}{3}{r_{\rm{max}}^3}+O\left(\frac{1}{r_{\rm{max}}}\right)\ .
\end{equation}
It is important that in these coordinates the divergent term is
independent of the field $\tilde l$, it is therefore possible to
regularize the on-shell action by subtracting the free energy of the
$\tilde l\equiv 0$ embedding. The resulting regularized expression for
the free energy is:
\begin{equation}
F=S_{\rm{reg}}=N_f\frac{\mu_5}{g_s}4\pi V_3R^3H^{3/2}\tilde I_{\rm{D5}}\ ,
\end{equation}
where
\begin{equation}
\tilde I_{\rm{D5}}=\int\limits_0^{\infty}d\tilde r\left[\tilde r^2\sqrt{1+\tilde l'^2}\sqrt{1+\frac{1}{(\tilde r^2+\tilde l^2)^2}} -\sqrt{1+\tilde r^4}\right]\, .
\end{equation}

\begin{figure}[h] 
   \centering
   \includegraphics[width=9cm]{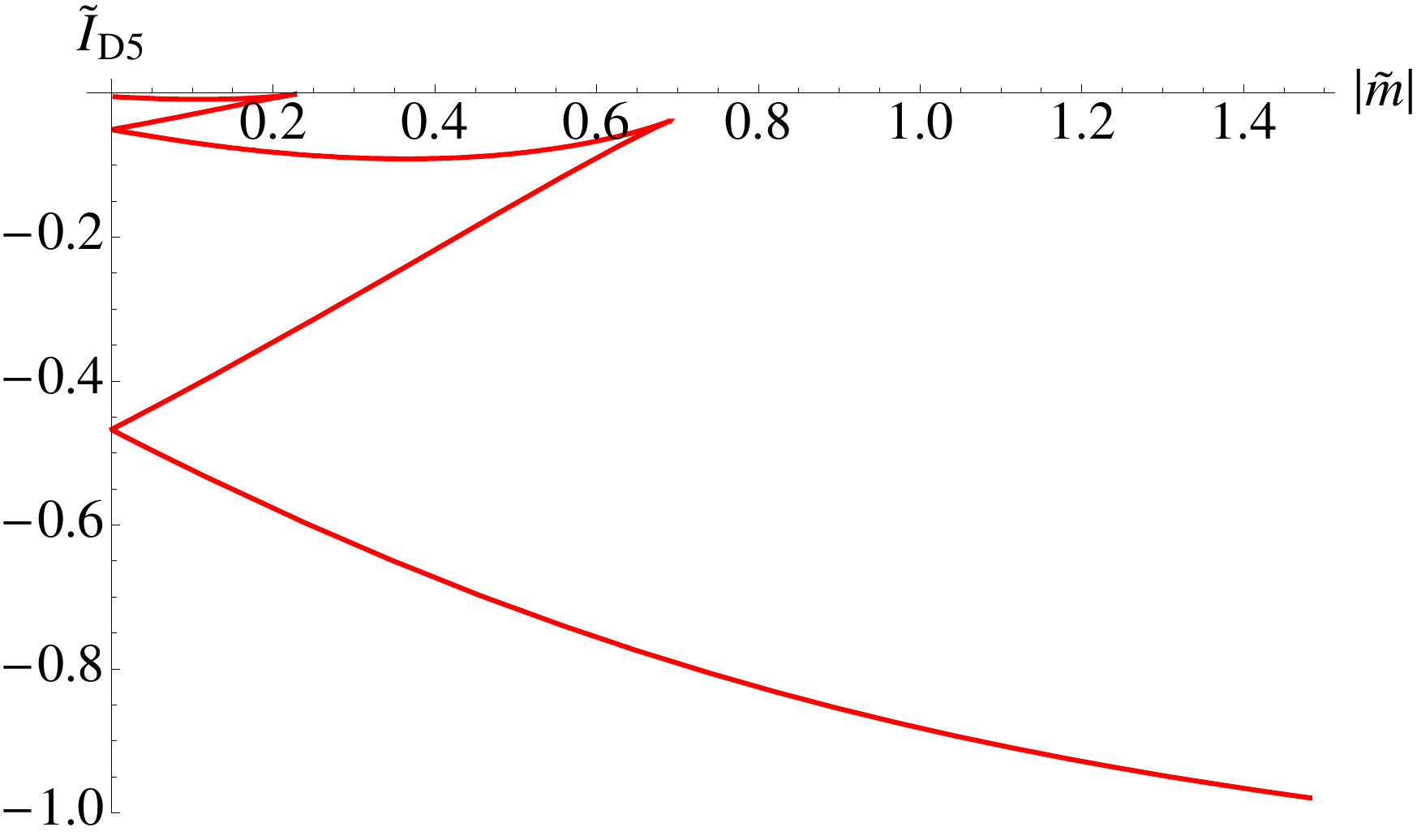}
   \caption{\small The states corresponding to the
     lowest positive branch of the plot in figure~\ref{fig:fig3} have
     the lowest free energy and thus correspond to the stable phase of
     the theory.  }
   \label{fig:fig4}
\end{figure}

\begin{figure}[h] 
   \centering
   \includegraphics[width=9cm]{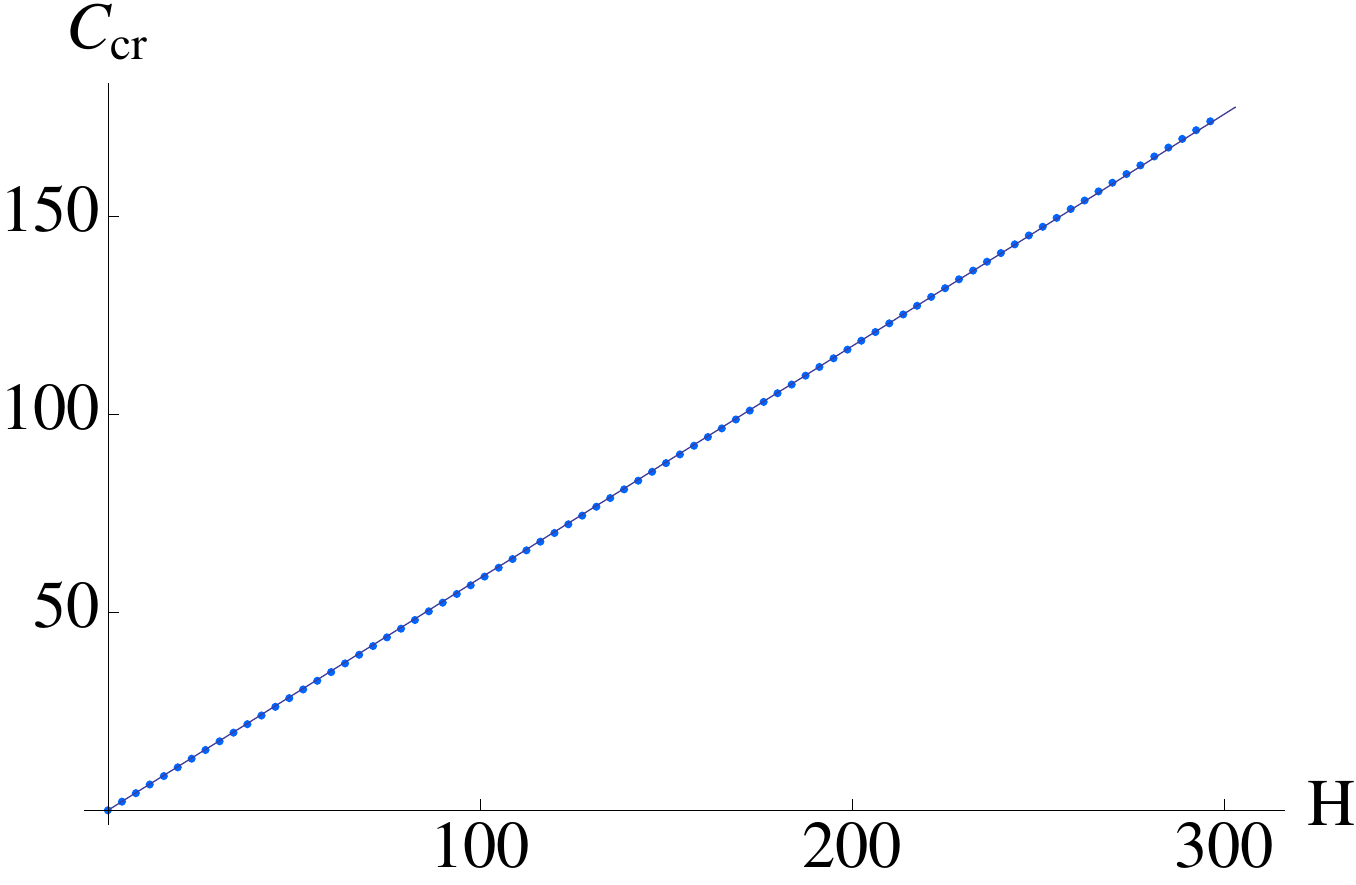}
   \caption{\small A plot of $c_{\rm{cr}}$ {\it vs.} $H$ for
     $R=1$. The solid (black) line corresponds to the linear fit
     $c_{\rm{cr}}\approx 0.586H$.  }
   \label{fig:fig5}
\end{figure}

A plot of $\tilde I_{\rm{D5}}$ {\it vs.} $|\tilde m|$ is presented in
figure~\ref{fig:fig4}. The states from the lowest
positive branch in figure~\ref{fig:fig4} have the lowest free energy
and correspond to the stable phase of the theory. Therefore there is a
spontaneous breaking of the global $SO(3)$ symmetry and the theory at
$\tilde m=0$ develops a negative condensate proportional to $-\tilde
c_{\rm{cr}}\approx -0.59$. Note that only the absolute value of
$\tilde m$ corresponds to the bare mass of the fundamental fields. The
states with negative $\tilde m$ correspond to D5--brane embeddings that
intercept the $\tilde l=0$ line in the $\tilde l$ {\it vs.} $\tilde r$
plane and as seen from figure~\ref{fig:fig4} are unstable. It is to be 
expected that the meson spectrum of the theory in such a phase would
contain tachyons based on an analogy with the meson spectrum of the
D3/D7 system studied in ref.~\cite{Filev:2007qu}.
Before we proceed with the analysis of the meson spectrum of the
theory let us write an expression for the condensate of the theory
$\langle\bar\psi\psi\rangle\propto -c_{\rm{cr}}=R^2H\tilde
c_{\rm{cr}}$ at zero bare quark mass. The coefficient of
proportionality is given by \cite{Mateos:2006nu}:
\begin{equation}
\langle\bar\psi\psi\rangle=-8\pi^2\alpha'\frac{\mu_5}{g_s}c_{\rm{cr}}=-16\pi^3\alpha'^2\frac{\mu_5}{g_s}\tilde c_{\rm{cr}}R^2(H/2\pi\alpha')\, .
\end{equation}
Note that the condensate is proportional to the magnitude of the
magnetic field $H/2\pi\alpha'$. In order to check the consistency of
our numerical analysis and to calculate more accurately the constant
$\tilde c_{\rm{cr}}$ we have calculated the value of $c_{\rm{cr}}$ for
a range of $H$ having set $R=1$. The resulting plot is presented in
figure~\ref{fig:fig5}. The solid (black) line corresponds to the
linear fit $c_{\rm{cr}}\approx 0.586H$ therefore we have $\tilde
c_{\rm{cr}}\approx
0.586$.

\subsection{Meson spectrum and pseudo-Golstone bosons}

In this section we will analyze the normal modes of the
D5--brane. These describe fluctuations of the spinor bilinear in the
dual gauge theory and hence their spectrum is the spectrum of the
light meson--like excitations of the gauge theory. We focus our
analysis on the normal modes corresponding to the Goldstone bosons
(which we label as pions here for simplicity) of the spontaneously
broken $SO(3)$ symmetry and study their spectrum as a function of the
bare quark mass $m_q$. Our study shows that the external magnetic
field splits the degeneracy of the meson spectrum and gives mass to
one of the pions of the theory. It also modifies the standard
$M_{\pi}^2\propto m$ GMOR relation for the remaining Goldstone mode to
a linear relation $M_{\pi}\propto m$. We will show that these results
are in accord with the behavior expected from the effective chiral
Lagrangian of the theory.

In order to study the light meson spectrum of the theory we look for
the quadratic fluctuations of the D5--brane embedding along the
transverse directions parametrized by $l,\psi,\phi$. To this end we
expand:
\begin{equation}
l=\bar l+2\pi\alpha'\delta l;~~~\psi=2\pi\alpha'\delta\psi;~~~\phi=2\pi\alpha'\delta\phi\, ,
\end{equation}
in the action (\ref{DBI2}) and leave only terms of order
$(2\pi\alpha')^2$. Note that fluctuations of the $U(1)$ gauge field
$F_{\alpha\beta}$ of the D5--brane will also contribute to the
expansion. There is also an additional contribution from the
Wess-Zumino term of the D5--brane's action:
\begin{equation}
S_{\rm{WZ}}=N_f\mu_5\int\limits_{{\cal M}_6}\sum_p[C_{p}\wedge e^{\cal F}];~~~{\cal F}=B+2\pi\alpha'F\, .
\end{equation}
For the anzatz that we are considering, the relevant term is:
\begin{equation}
S_{\rm{WZ}}=N_f\mu_5\int\limits_{{\cal M}_6}B\wedge P[\tilde C_{4}]\ ,
\end{equation}
where $P[\tilde C_4]$ is the pull-back of the magnetic dual, $\tilde
C_{4}$, to the background $C_{4}$ R-R form. For the particular
parameterization of $S_5$ considered here, it is given by:
\begin{equation}
\tilde C_4=\frac{1}{g_s}\frac{4r^2l^2}{(r^2+l^2)^3}R^4\sin\psi(l dr- rdl)\wedge d\Omega_2\wedge d\phi\ .
\end{equation}
After some long but straightforward calculations we get the following
action for the quadratic fluctuations along $l$:
\begin{eqnarray}
  &&{\cal L}_{ll}^{(2)}\propto\frac{1}{2}\sqrt{-E}G_{ll}\frac{S^{\alpha\beta}}{1+l'^2}\partial_{\alpha}\delta l\partial_{\beta}\delta l+\frac{1}{2}\left[\partial_{l}^2\sqrt{-E}-\frac{d}{dr}\left(\frac{l'}{1+l'^2}\partial_l\sqrt{-E}\right)\right]\delta l^2\ ,\\ 
  &&{\cal L}_{lF}^{(2)}\propto\frac{\sqrt{-E}}{1+l'^2}(\partial_l J^{12}-\partial_{r}J^{12}l')F_{21}\delta l\nonumber\ ,\\
  &&{\cal L}_{FF}^{(2)}\propto\frac{1}{4}\sqrt{-E}S^{\alpha\beta}S^{\gamma\lambda}F_{\beta\gamma}F_{\alpha\lambda}\nonumber\ ,
\end{eqnarray}
and along $\phi$ and $\psi$:
\begin{eqnarray}
&&{\cal L}_{\psi\psi,\phi\phi}^{(2)}\propto\frac{1}{2}\sqrt{-E}S^{\alpha\beta}(G_{\psi\psi}\partial_{\alpha}\delta\psi\partial_{\beta}\delta\psi+G_{\phi\phi}\partial_{\alpha}\delta\phi\partial_{\beta}\delta\phi)\ \label{qvfl},\\
&&{\cal L}_{\psi\phi}^{(2)}\propto (\cos\alpha) PH\delta\psi\partial_0\delta\phi\nonumber \ .
\end{eqnarray}
Here $E_{\alpha\beta}$ is the pull-back of the generalized metric on
the classical D5--brane embedding:
\begin{equation}
E_{\alpha\beta}=\partial_{\alpha}\bar X^{\mu}\partial_{\beta}\bar X^{\nu}(G_{\mu\nu}+B_{\mu\nu})\, ,
\end{equation}
and we have defined $S^{\alpha\beta}$ and $J^{\alpha\beta}$ as the
symmetric and anti-symmetric elements of the inverse generalized
metric $E^{\alpha\beta}$:
\begin{equation}
E^{\alpha\beta}=S^{\alpha\beta}+J^{\alpha\beta}\ .
\end{equation}
The determinant $E$ and the function $K=P$ are given by:
\begin{eqnarray}
&&\sqrt{-E}=(\cos\alpha) r^2\sqrt{1+l'^2}\sqrt{1+\frac{R^4H^2}{(r^2+l^2)^2}}\equiv g(r)\cos\alpha \ ,\\
&&P=\frac{4R^4r^2l^2}{(r^2+l^2)^3}(rl'-l)\ .
\end{eqnarray}
As one can see, the fluctuations along $\psi$ and $\phi$ decouple from
the fluctuations along $l$ and the fluctuations of the gauge field
$A_{\alpha}$. Since we are interested in the pseudo-Goldstone modes of
the dual theory we will focus on the fluctuations along $\psi$ and
$\phi$. The equations of motion derived from the quadratic action
(\ref{qvfl}) are the following:
\begin{eqnarray}
&&\partial_r\left(\frac{g(r)l^2}{1+l'^2}\partial_r\delta\psi\right)+\frac{g(r)R^4l^2}{(r^2+l^2)^2}\tilde\Box\delta\psi+\frac{g(r)R^4l^2r^2}{(r^2+l^2)^2}\Delta_{(2)}\delta\psi-PH\partial_0\delta\phi=0\ ,\\
&&\partial_r\left(\frac{g(r)l^2}{1+l'^2}\partial_r\delta\phi\right)+\frac{g(r)R^4l^2}{(r^2+l^2)^2}\tilde\Box\delta\phi+\frac{g(r)R^4l^2r^2}{(r^2+l^2)^2}\Delta_{(2)}\delta\phi+PH\partial_0\delta\psi=0\ ,\nonumber
\end{eqnarray}
where 
\begin{equation}
\tilde\Box=-\partial_0^2+\frac{\partial_1^2+\partial_2^2}{1+\frac{R^4H^2}{(r^2+l^2)^2}}\ .
\label{LP}
\end{equation}
Note that the background magnetic field breaks the $SO(1,2)$ Lorentz
symmetry to $SO(2)$, which manifests itself in the modified laplacian
(\ref{LP}). Next we consider a plane-wave ansatz:
\begin{equation}
\delta\phi=e^{i(\omega t-\vec k\dot \vec x)}\eta_1(r);~~~\delta\psi=e^{i(\omega t-\vec k\dot \vec x)}\eta_2(r)\label{anzatsqv}\, ,
\end{equation}
now using the anzatz (\ref{anzatsqv}) we get:
\begin{eqnarray}
&&\partial_r\left(\frac{g(r)l^2}{1+l'^2}\eta_1'\right)+\frac{g(r)R^4l^2}{(r^2+l^2)^2}(\omega^2-\frac{{\vec k}^2}{1+\frac{R^4H^2}{(r^2+l^2)^2}})\eta_1-i\omega PH\eta_2=0\ ,\label{coupled}\\
&&\partial_r\left(\frac{g(r)l^2}{1+l'^2}\eta_2'\right)+\frac{g(r)R^4l^2}{(r^2+l^2)^2}(\omega^2-\frac{{\vec k}^2}{1+\frac{R^4H^2}{(r^2+l^2)^2}})\eta_2+i\omega PH\eta_1=0\ .\nonumber
\end{eqnarray}
The equations of motion in (\ref{coupled}) can be decoupled by the
definition $\eta_{\pm}=\eta_1\pm i\eta_2$. The result is:
\begin{eqnarray}
&&\partial_r\left(\frac{g(r)l^2}{1+l'^2}\eta_+'\right)+\frac{g(r)R^4l^2}{(r^2+l^2)^2}(\omega^2-\frac{{\vec k}^2}{1+\frac{R^4H^2}{(r^2+l^2)^2}})\eta_+-\omega PH\eta_+=0\ ,\label{decoupled}\\
&&\partial_r\left(\frac{g(r)l^2}{1+l'^2}\eta_-'\right)+\frac{g(r)R^4l^2}{(r^2+l^2)^2}(\omega^2-\frac{{\vec k}^2}{1+\frac{R^4H^2}{(r^2+l^2)^2}})\eta_-+\omega PH\eta_-=0\ .\nonumber
\end{eqnarray}
Because of the broken Lorentz symmetry, the $1+2$ dimensional mass
$M^2=\omega^2-{\vec k}^2$ depends on the choice of frame. We can
define the spectrum of excitations as the rest energy (consider the
frame with $\vec k=0$) and as we shall observe, the spectrum is
discrete. Furthermore just as in the D3/D7 case there is a Zeeman
splitting of the spectrum due to the external magnetic
field. Interestingly, at low energy the splitting is breaking the
degeneracy of the lowest energy state and as a result there is only
one pseudo-Goldstone boson. Note that this is not in contradiction
with the Goldstone theorem because there is no Lorentz symmetry. This
opens the possibility of having two types of Goldstone modes: type I
and type II satisfying odd and even dispersion relations
correspondingly. In this case there is a modified counting rule
(ref.~\cite{Nielsen:1975hm}, see also ref.~\cite{Brauner:2005di})
which states that {\it the number of GBs of type I plus twice the
  number of GBs of type II is greater than or equal to the number of
  broken generators.} As we are going to show below the single
Goldstone mode that we see satisfies a quadratic dispersion relation
(hence is type II) and the modified counting rule is not
violated. Note also that for large bare masses $m$ (and
correspondingly large values of $l$) the term proportional to the
magnetic field is suppressed and the meson spectrum should approximate
to the result for the pure AdS$_5\times S^5$ space-time case studied
in refs.~\cite{{Arean:2006pk},{Myers:2006qr}}, where the authors
obtained the following relation:
\begin{equation}
\omega_n=\frac{2m}{R^2}\sqrt{(n+1/2)(n+3/2)}\ ,\label{purespect}
\end{equation}
between the eigenvalue of the $n^{th}$ excited state $\omega_n$ and the bare mass $m$.  

\begin{figure}[h] 
   \centering
   \includegraphics[width=10cm]{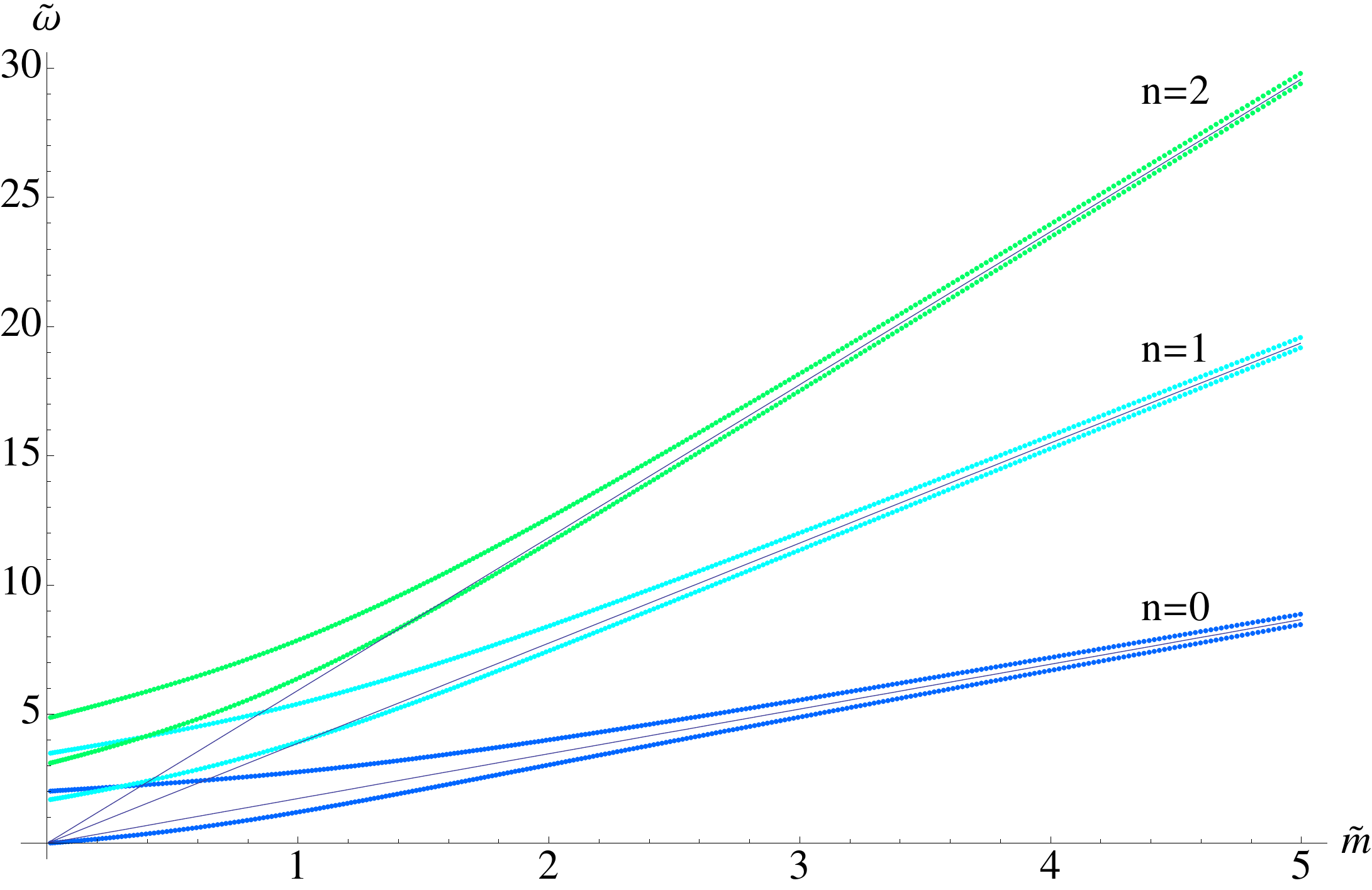}
   \caption{\small The meson spectrum of the first three excited
     states is plotted. There is Zeeman splitting of the
     spectrum and the existence of a mass gap at $\tilde m=0$ as well
     as a single Goldstone boson mode. For large $\tilde m$ the
     spectrum asymptotes to that of zero magnetic field given by
     equation (\ref{purespect}) (straight lines).}
   \label{fig:fig6}
\end{figure}

\begin{figure}[h] 
   \centering
   \includegraphics[width=10cm]{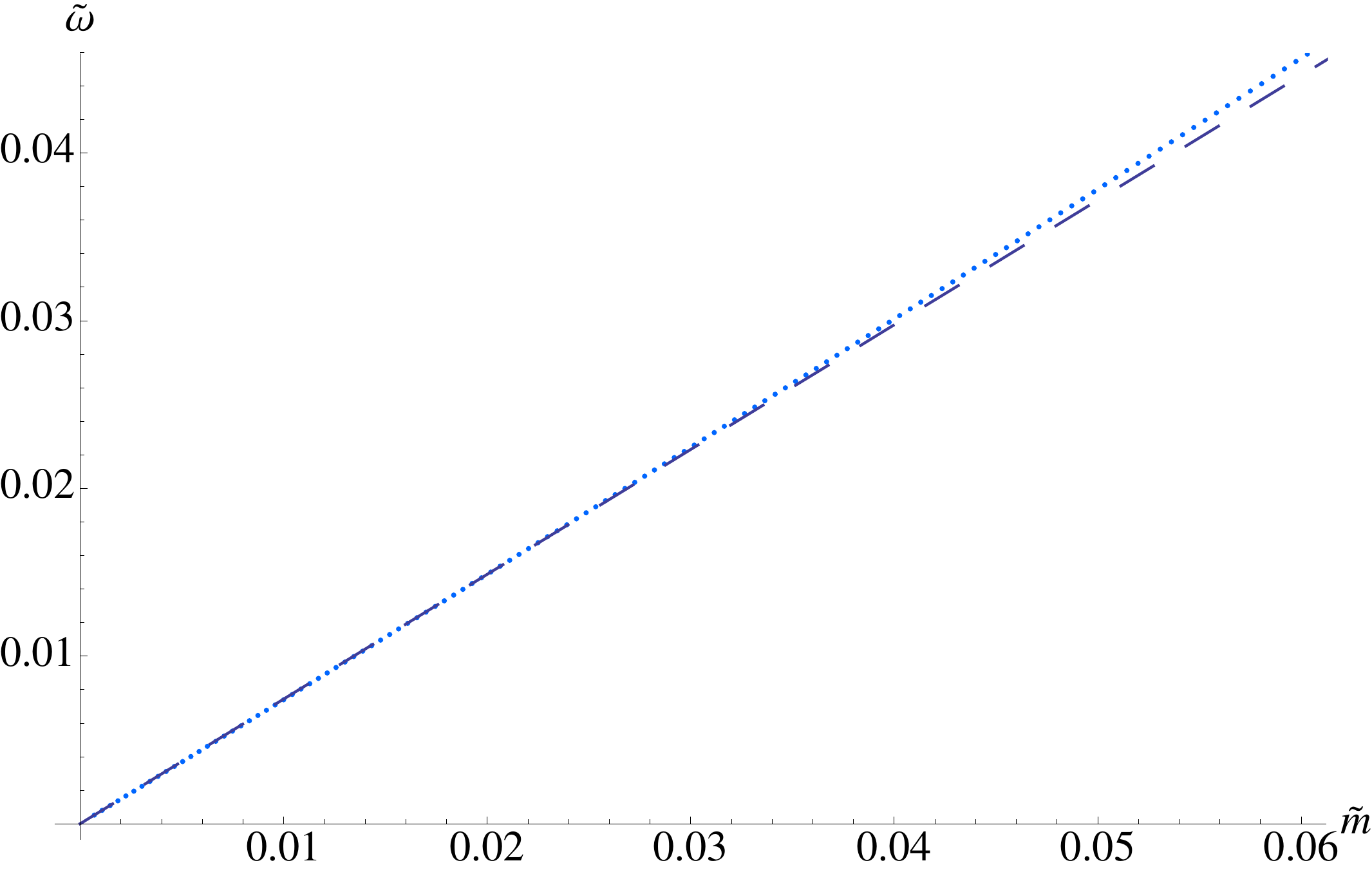}
   \caption{\small Plot of the spectrum of the ground state from
     figure~\ref{fig:fig6} for small bare masses. The dashed line
     corresponds to the linear behavior from equation
     (\ref{lingmor}).}
\label{fig:fig7}
\end{figure}

In order to obtain the meson spectrum, we numerically solve the
equations of motion (\ref{decoupled}) in the rest frame ($\vec
k=0$). The quantization condition for the spectrum comes from imposing
regularity at infinity. More precisely we require that $\eta_{\pm}
\sim 1/r$ at infinity ($r\to\infty$). The results are summarized in
figure~\ref{fig:fig6}. Just as in the D3/D7 case we have defined the
dimensionless quantities $\tilde m=m/R\sqrt{H}$ and
$\tilde\omega=\omega R/\sqrt{H}$. As one can see from figure
\ref{fig:fig6}, for large $\tilde m$ the spectrum asymptotes to that
of pure AdS$_5\times S^5$, given by equation~(\ref{purespect}). The
Zeeman splitting of the spectrum is also evident. It is interesting
that as a result of the splitting of the ground state there is only a
single pseudo-Goldstone mode. Furthermore, as can be seen from figure
\ref{fig:fig7}, for small bare masses instead of the usual
Gell-Mann--Oakes--Renner relation we obtain a linear dependence
$\tilde\omega\sim\tilde m$. As we will show in the next subsection the
slope is given by the relation:
\begin{equation}
\tilde\omega=\frac{4\tilde c_{\rm{cr}} }{\pi}\tilde m\approx 0.736\tilde m\, . \label{lingmor}
\end{equation}

It is also interesting to study the dispersion relation of the
Goldstone mode. Since we have broken Lorentz symmetry and observe only
one pseudo-Goldstone mode (which is only half the number of broken
generators) we anticipate a quadratic dispersion relation (see
refs.~\cite{Brauner:2005di} and~\cite{Brauner:2007uw} for discussion).

\begin{figure}[h] 
   \centering
   \includegraphics[width=10cm]{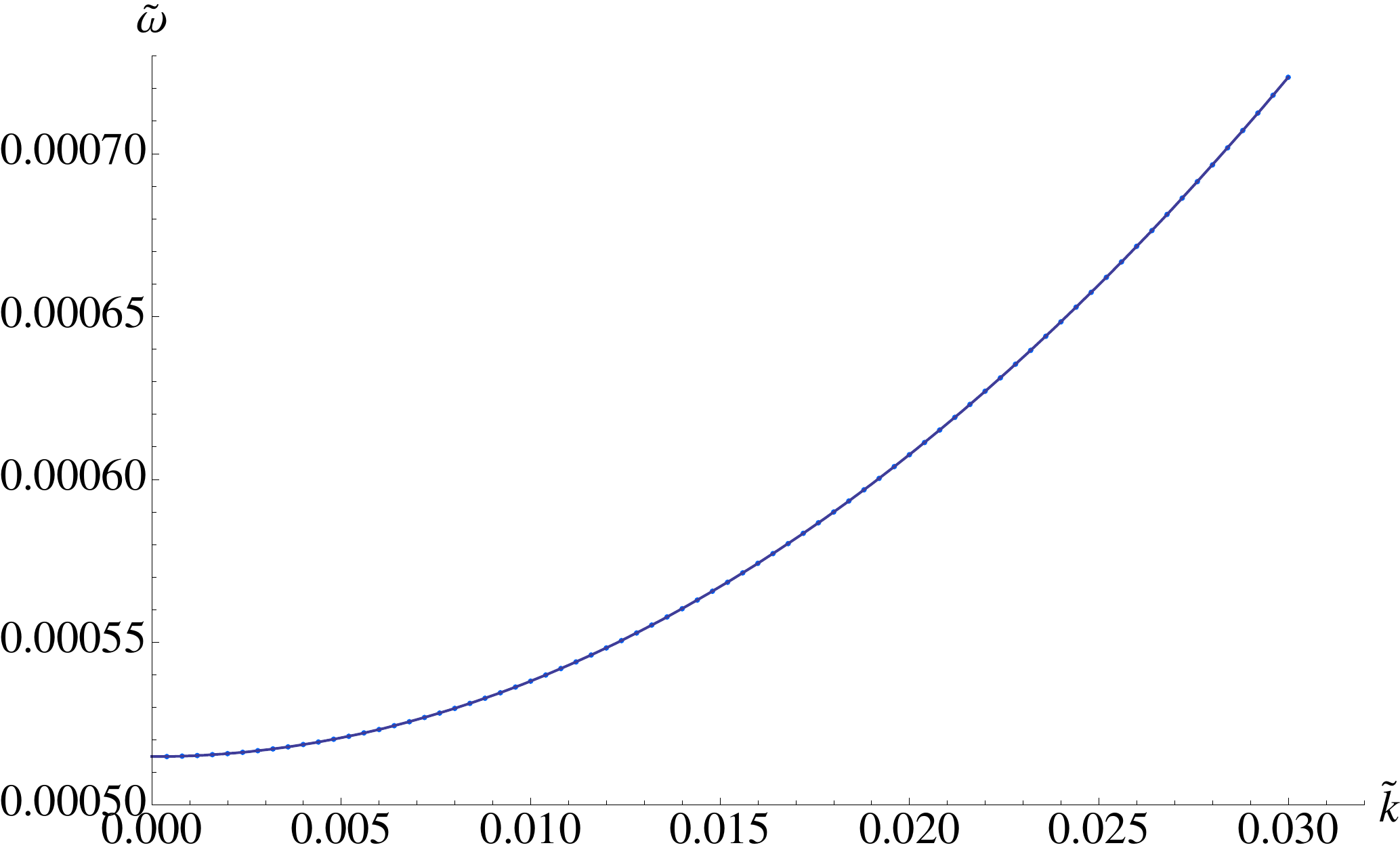}
   \caption{\small Plot of the  dispersion relation of the pseudo-Goldstone mode for $\tilde m\approx 0.0007$. The parabolic fit corresponds to equation (\ref{dispfit}).}
\label{fig:fig8}
\end{figure}

In order to obtain the dispersion relation of the Goldstone mode we
numerically solve equations (\ref{decoupled}) at very small bare mass
$\tilde m\approx 0.0007$ and for a range of small momenta $\tilde{\vec
  k}=\vec k R/\sqrt{H}$. The result is presented in figure
\ref{fig:fig8}. There is  indeed  a quadratic dispersion
relation. As we are going to show, the dispersion relation is given
by:
\begin{equation}
\tilde \omega=\gamma {\tilde{\vec k}}^2+\frac{4}{\pi}\tilde c_{\rm{cr}}\tilde m\label{dispersion}\ ,
\end{equation}
where:
\begin{equation}
\gamma=\frac{4}{\pi}\int\limits_{0}^{\infty}d\tilde r\frac{\tilde r^2\tilde l^2\sqrt{1+\tilde l'^2}}{(\tilde r^2+\tilde l^2)\sqrt{1+(\tilde r^2+\tilde l^2)^2}}\, .\label{gamma}
\end{equation}
For $\tilde m\approx 0.0007$ the relation (\ref{dispersion}) is given by:
\begin{equation}
\tilde \omega\approx 0.232 {\tilde{\vec k}}^2+0.000515\, ,\label{dispfit}
\end{equation}
and is represented by the fitted curve in figure~\ref{fig:fig8}.

In the next subsections we will obtain the effective $1+2$ dimensional
chiral action describing the pseudo-Goldstone mode and argue that in
the limit $\omega\to 0$ it is identical to the action describing
magnon excitations in a ferromagnet \cite{Hofmann:1998pp}. Furthermore
we will show that the observed dispersion relation (\ref{dispersion})
is in agreement with the dispersion relation of magnons in an external
magnetic field. Note that in order to make the analogy with a
ferromagnet, one needs to identify the bare mass with the external
magnetic field acting on the ferromagnet. The reason is that these
both correspond to the small parameter that explicitly breaks the global symmetry.

\subsubsection{Low energy dispersion relation}  
In order to obtain the dispersion relation for the pseudo-Goldstone
mode we will analyze the first equation in (\ref{decoupled}) in the
spirit of the analysis performed in section 2.2.1 for the D3/D7
system. To begin with let us consider the limit of small $\omega$ thus
leaving only the linear potential term in $\tilde \omega$. In view of
the observed quadratic dispersion relation (\ref{dispersion}) we will
also keep the ${\vec k}^2$ term in equation~(\ref{decoupled}).
\begin{equation}
\partial_r\left(\frac{g(r)l^2}{1+l'^2}\eta_+'\right)-\left(\omega PH+\frac{g(r)R^4l^2}{(r^2+l^2)^2+R^4H^2}{\vec k}^2\right)\eta_+=0\, .\label{etapl}
\end{equation}
It is convenient to define the following variables:
\begin{equation}
\Theta^2=\frac{g(r)l^2}{1+l'^2}\ ;~~~\xi=\eta_+\Theta\ .
\end{equation}
Then equation~(\ref{etapl}) can be written as:
\begin{equation}
\ddot\xi-\frac{\ddot\Theta}{\Theta}\xi-\left(\omega PH+\frac{g(r)R^4l^2}{(r^2+l^2)^2+R^4H^2}{\vec k}^2\right)\frac{\xi}{\Theta^2}=0\ .\label{xi}
\end{equation}
Where the overdots represent derivatives with respect to $r$. Now if we take
the limit $m\to 0$ we have that $\omega\to 0$ and $k\to 0$ and obtain
that:
\begin{equation}
\xi=\Theta|_{\omega=0}\equiv\bar\Theta\ ,
\end{equation}
is a solution to equation (\ref{xi}). Our next step is to consider small $m$ and expand:
\begin{equation}
\xi=\bar\Theta+\delta\xi\ ;~~~\Theta=\bar\Theta+\delta\Theta\ ,\label{expcrit}
\end{equation}
where the variations $\delta\xi$ and $\delta\Theta$ are vanishing in
the $m\to 0$ limit. Then, to leading order in $m$ (keeping in mind
that $\omega\sim m$ and $\vec k^2\sim m$) we obtain:
\begin{equation}
\delta\ddot\xi-\frac{\ddot{\bar\Theta}}{\bar\Theta}\delta\xi-\delta\left(\frac{\ddot\Theta}{\Theta}\right)\bar\Theta-\left(\omega PH+\frac{g(r)R^4l^2}{(r^2+l^2)^2+R^4H^2}{\vec k}^2\right)\frac{1}{\bar\Theta}=0\ .
\label{prib}
\end{equation}
Now we multiply equation (\ref{prib}) by $\bar\Theta$ and integrate
along $r$. The result is:
\begin{equation}
(\bar\Theta\delta\dot\xi-\dot{\bar\Theta}\delta\xi)\Big |_0^{\infty}-(\bar\Theta\delta{\dot\Theta}-\dot{\bar\Theta}\delta\Theta)\Big |_0^{\infty}-\omega H\int\limits_0^{\infty}{dr}P(r)-\frac{\pi}{4} R^5\sqrt{H}\gamma\vec k^2=0\ .\label{intermid}
\end{equation}
Using the definitions of $\Theta, P(r)$ and $\xi$ and requiring regularity at infinity for $\eta_+$, one can show that the first term in equation~(\ref{intermid}) vanishes and that:
\begin{equation}
 (\bar\Theta\delta{\dot\Theta}-\dot{\bar\Theta}\delta\Theta)\Big |_0^{\infty}=c\delta  m\ ;~~~\int\limits_0^{\infty}{dr}P(r)=-R^4\pi/4\ ,
\end{equation}
and hence using the previous definitions, $\tilde m =m/R\sqrt{H}$, $\tilde c=c/R^2 H$, $\tilde\omega=\omega R/\sqrt{H}$ and $\tilde{\vec k}=\vec k R/\sqrt{H}$, we obtain equation (\ref{dispersion}) which we duplicate below:
\begin{equation}
\tilde \omega=\gamma {\tilde{\vec k}}^2+\frac{4}{\pi}\tilde c_{\rm{cr}}\tilde m\ .\label{disp2}
\end{equation}
In the next subsection we will derive the effective $1+2$ dimensional
action for the pseudo-Goldstone mode and show that to leading order it
is in to one correspondence with the effective action describing
magnon excitations in a ferromagnet corresponding to the $SO(3)\to
SO(2)$ spontaneous symmetry breaking by spontaneous magnetization
\cite{Hofmann:1998pp}. We will relate the fermionic condensate $\tilde
c$ to the spontaneous magnetization of the ferromagnet and the bare
mass to the external magnetic field and show that the dispersion
relation (\ref{disp2}) is in exact agreement with that of magnons.

\subsubsection{Effective chiral Lagrangian}

In order to obtain the $1+2$ dimensional effective action describing
the pseudo-Goldstone mode we consider the $1+5$ dimensional action
(\ref{qvfl}) for a classical embedding in the vicinity of the critical
embedding, namely that embedding corresponding to a very small bare
mass $\tilde m$. Now let us consider the following ans\"atze for the
fields $\delta\phi$ and $\delta\psi$:
 \begin{equation} 
\delta\phi=\frac{\xi_1(r)}{\Theta(r)}\chi_1(x)\ ;~~~\delta\psi=\frac{\xi_2(r)}{\Theta(r)}\chi_2(x)\ .
\end{equation}
Since we are close to the critical embedding we will consider the same expansion as in equation~(\ref{expcrit}):
\begin{equation}
\xi_i=\bar\Theta+\delta\xi_i\ ,~~~i=1\, {\rm or}\, 2\ ;~~~\Theta=\bar\Theta+\delta\Theta\ .
\end{equation}
By definition it follows that as $\tilde m\to 0$, $\delta \xi_i$ and $\delta \Theta$ vanish. Then to leading order we have that:
\begin{eqnarray}
&&\partial_r\delta\phi=\frac{1}{\bar\Theta^2}[(\bar\Theta\delta\dot\xi_1-\dot{\bar\Theta}\delta\xi_1)+(\dot{\bar\Theta}\delta\Theta-\bar\Theta\delta\dot\Theta)]\chi_1(x)\ ;~~~\partial_{\mu}\delta\phi=\partial_{\mu}\chi_1(x)\ ;~~\mu=0,1,2\, ,\\
&&\partial_r\delta\psi=\frac{1}{\bar\Theta^2}[(\bar\Theta\delta\dot\xi_2-\dot{\bar\Theta}\delta\xi_2)+(\dot{\bar\Theta}\delta\Theta-\bar\Theta\delta\dot\Theta)]\chi_2t(x)\ ;~~~\partial_{\mu}\delta\psi=\partial_{\mu}\chi_2(x)\ ;~~\mu=0,1,2\, .\nonumber
\end{eqnarray}
Now we integrate equation~(\ref{qvfl}) along $r$ from $0,\infty$ and
along the internal unit sphere $\tilde\Omega_2$. The interesting term
is the part of the kinetic term involving derivatives along $r$. After
integration by parts it boils down to a mass term for the $1+2$
dimensional fields $\chi_1,\chi_2$. Explicitly:
\begin{eqnarray}
&&\int drd\tilde\Omega_2\frac{1}{2}\frac{g(r)l^2}{1+l'^2}\partial_r\delta\phi\partial_r\delta\phi=-\int drd\tilde\Omega_2\frac{1}{2}\partial_r\left(\frac{g(r)l^2}{1+l'^2}\partial_r\delta\phi\right)\delta\phi=\label{kinrph}\\
&&\hskip2cm =-4\pi[(\bar\Theta\delta\dot\xi_1-\dot{\bar\Theta}\delta\xi_1)+(\dot{\bar\Theta}\delta\Theta-\bar\Theta\delta\dot\Theta)]\Big|_0^{\infty}\frac{1}{2}\chi_1^2=4\pi m c\frac{1}{2}\chi_1^2\nonumber\ .
\end{eqnarray}
Here we have used the same arguments as in equation
(\ref{intermid}). It is clear that one can perform an analogous
calculation for the term involving $\partial_r\delta\psi$. The rest of
the terms are dealt with straightforwardly by integrating along
$r$. The resulting action is:
\begin{equation}
\frac{S_{\rm{eff}}}{(2\pi\alpha')^2}=\int d^3x\left\{\frac{f_{\pi||}^2}{4}\partial_0\chi^{*}\partial_0\chi-\frac{f_{\pi\bot}^2}{4}\partial_i\chi^{*}\partial_i\chi-\mu\frac{i}{2}(\chi\partial_0\chi^{*}-\chi^{*}\partial_0\chi)+\frac{m_q}{2}\langle\bar\Psi\Psi{\rangle}_0\chi^*\chi\right\}\ ,\label{effectivef}
\end{equation}
where we have defined a complex scalar field
$\chi=\chi_1+i\chi_2$. The constants in the effective action are given
by:
\begin{eqnarray}
&&\frac{f_{\pi||}^2}{4}=\frac{{\cal N}}{2}\int\limits_0^{\infty} dr\frac{g(r)R^4l^2}{(r^2+l^2)^2}\ ;~~~~\frac{f_{\pi\bot}^2}{4}=\frac{{\cal N}}{2}\int\limits_0^{\infty} dr\frac{g(r)R^4l^2}{(r^2+l^2)^2+R^4H^2}\ ,\\
&&\mu=\frac{\cal N}{8}\pi R^4H;~~\langle\bar\psi\psi\rangle=-(2\pi\alpha'){\cal N}c_{\rm{cr}}\ ;~~{\cal N}=4\pi N_f\frac{\mu_5}{g_s};~~m_q=\frac{m}{2\pi\alpha'}\ .\nonumber
\end{eqnarray}
The effective action (\ref{effectivef}) is very
similar to the one considered in ref.~\cite{Brauner:2005di}, where the
author studied Goldstone bosons in linear sigma models with chemical
potential, only we have further broken the Lorentz symmetry by the
introduction of an external magnetic field. The peculiar feature of
this effective Lagrangian is the single time derivative term that is
responsible for the unusual quadratic dispersion relation. In
ref. \cite{Brauner:2005di} the authors have shown that the number of
goldstone modes with quadratic dispersion relation is half the number
of broken generators. This is exactly what we observe here (two broken
generators but only a single goldstone mode).

On the other hand, for the pseudo-Goldstone mode $\omega\to 0$ and to
leading order the effective action (\ref{effectivef}) can be written
as:
\begin{equation}   
S_{\rm{eff}}^{f}=\int d^3x\left\{\frac{1}{2}\Sigma\epsilon_{ab}\partial_0U^aU^b-\Sigma\beta h\frac{1}{2}{U^a}^2-\frac{1}{2}F^2\partial_iU^a\partial_iU^a\right\}\ ;~~~a=1\,{\rm or}\, 2\ ,\label{fero}
\end{equation}
where
\begin{equation}
U^a=(2\pi\alpha')\chi^a\ ;~~~\Sigma=2\mu;~~~F^2=\frac{f_{\pi\bot}^2}{2}\ ;~~~\beta h=-\frac{m_q\langle\bar\psi\psi\rangle}{2\mu}=\tilde c_{\rm{cr}}\tilde m\frac{4\sqrt{H}}{\pi R}\ .\label{condmat}
\end{equation}
Equation (\ref{fero}) corresponds to the effective action describing a
ferromagnet in an external magnetic field $h$ with a spontaneous
magnetization $\Sigma$ and a magnetic coupling $\beta$
\cite{Hofmann:1998pp}. Here $\vec U=(U^1,U^2,U^3)$ is a unit vector
corresponding to the direction of the spontaneous magnetization of the
ferromagnet, and the action (\ref{fero}) is describing quadratic
fluctuations of the magnetization near the ordered state $U^3=1$. The
fact that the effective external magnetic field $h$ is proportional to
the bare mass $\tilde m$ is to be expected since in both descriptions
these are the small parameters that reduce the exact global symmetry to an approximate one by coupling to the corresponding order parameters (magnetization and quark condensate corresponding). If one takes into account the re-definitions from equation (\ref{condmat}) and the
definitions of $\tilde\omega$ and $\tilde{\vec k}$, it is
straightforward to check that the dispersion relation of ferromagnetic
spin waves \cite{Hofmann:1998pp}:
\begin{equation}
\omega=\gamma k^2+\beta h\ ;~~~\gamma=\frac{F^2}{\Sigma}\ ,
\end{equation}
is exactly that of equation (\ref{dispersion}). 

Of course many different microscopic systems exhibit the same low
energy behavior and hence are described by the same effective
Lagrangian. Furthermore the fact that the mass generation process in
the D3/D5 system is associated to precisely the same global symmetry
breaking pattern ($SO(3)\to SO(2)$) as the transition from
paramagnetic to ferromagnetic phase is also very suggestive. However
it is the peculiar single derivative term coming from the Wess-Zumino
contribution that is responsible for the observed dispersion
relation. Here we have not investigated how far one can go in
describing properties of ferromagnets using the D3/D5 set up. It would
be interesting to study interaction terms, which would require
expanding the effective action beyond quadratic order. In any case it
is somewhat satisfying that the dispersion relation of
pseudo--Goldstone modes can be related to a real condensed matter
phenomenon such as magnon spin waves.

\section{Universal properties of magnetic catalysis in  Dp/Dq systems}

In this section we focus on some universal features of the mechanism
of spontaneous symmetry breaking in an external magnetic field in the
context of the Dp/Dq system. In particular we explore the observed
discrete self--similar behavior of the theory in the vicinity of the
trivial embedding $l\equiv 0$ corresponding to the non-symmetry
breaking phase. As we mentioned in Section 2 and Section 3, for the
D3/D7 and the D3/D5 systems, this embedding is unstable and the
instability is manifested as a multi-valuedness of the equation of
state in the condensate versus bare mass plane $(\tilde m,\tilde c)$
seeded by a logarithmic spiral structure (see figure~\ref{fig:fig2}
and figure~\ref{fig:fig3}). This spiral structure has been explored in
details in ref.~\cite{Filev:2007gb} in the case of the D3/D7 set
up. It has been shown that the spectrum of meson--like excitations also
exhibits discrete self--similar structure in the tachyonic sector of
the theory. It is interesting that the same structure appears for
other first order phase transition in the Dp/Dq set up, such as the
meson melting \cite{Mateos:2006nu} and electrically driven
insulator/conductor phase transitions \cite{Albash:2007bq},
\cite{Filev:2008xt}. In ref.~\cite{Mateos:2006nu} it was pointed out
that the critical exponents (or more appropriately ``scaling
exponents") characterizing the logarithmic structure exhibit universal
properties and depend only on the dimension of the internal $S^n$
sphere wrapped by the Dq--brane. A similar analysis was performed in
ref.~\cite{Filev:2008xt} for the case of electrically and R-charge
chemical potential driven phase transitions and another set of
``scaling exponents" was obtained. Here we will extend the analysis of
the spiral structure in the case of the D3/D7 system performed in
ref.~\cite{Filev:2007gb} to the general case of Dp/Dq systems T-dual
to the D3/D7 intersection and will show that the corresponding scaling
exponents guarantee the existence of a discrete self--similar behavior
in all Dp/Dq systems of potential phenomenological interest. This
suggests the universal role of the external magnetic field as a strong
catalyst of mass generation.

To begin with, let us consider the zero temperature Dp--brane solution,
given by:
\begin{eqnarray}\label{S4:1}
&&ds^2=K_p^{-\frac{1}{2}}\left(-dt^2+\sum\limits_{i=1}^{p}dx_i^2\right)+K_p^{\frac{1}{2}}\left({du^2}+u^2d\Omega_{8-p}^2\right)\ , \\
&&e^{\Phi}=g_s K_p^{(3-p)/4}\ ;~~~C_{01\dots p}=K_p^{-1}\ ,\nonumber
\end{eqnarray}
where $K_p(u)=(R/u)^{7-p}$ and $R$ is a length scale (the AdS radius
in the $p=3$ case). Now if we introduce a D$q$--brane probe having $d$
common space--like directions with the D$p$--brane, wrapping an
internal $S^n\subset S^{8-p}$ and extended along the holographic
coordinate $u$, we will introduce fundamental matter to the dual gauge
theory that propagates along a $(d+1)$--dimensional defect.

Next we parameterize the transverse $9-p$ plane $du^2+u^2d\Omega_{8-p}^2$ by:
\begin{equation}
d\rho^2+dL^2+\rho^2d\Omega_n^2+L^2d\Omega_{7-p-n}^2\ ,
\end{equation}
where $d\Omega_m^2$ is the metric on a unit radius $m$--sphere and
$\rho^2+L^2=u^2$. We also introduce an external magnetic field
$H/2\pi\alpha'$, corresponding to the $F_{p-1,p}$ component of the
field strength tensor, by fixing a constant $B$-field in the
$(x_{p-1},x_{p})$ plane:
\begin{equation}
B_{(2)}=Hdx_{p-1}\wedge dx_{p}\ .
\end{equation}
Then the DBI part of the Lagrangian governing the classical embedding of
the probe is given by\footnote{We consider only systems T--dual to the D3/D7 one, which imposes the constraint $p-d+n+1=4$.}:
\begin{equation}
{\cal L}\propto e^{-\Phi}\sqrt{-|g_{\alpha\beta}|}=\frac{\sqrt{|\Omega_n|}}{g_s}\rho^n\sqrt{1+L'^2}\sqrt{1+\frac{H^2R^{7-p}}{(\rho^2+L^2)^{\frac{7-p}{2}}}}\ .
\end{equation}
The equation of motion for the classical D$q$--brane embedding is given by:
\begin{equation}
\partial_{\rho}\left(\frac{\rho^nL'}{\sqrt{1+L'^2}}\sqrt{1+\frac{H^2R^{7-p}}{(\rho^2+L^2)^{\frac{7-p}{2}}}}\right)+\frac{7-p}{2}\frac{\rho^n\sqrt{1+L'^2}}{(\rho^2+L^2)^{\frac{9-p}{2}}}\frac{LH^2R^{7-p}}{\sqrt{1+\frac{H^2R^{7-p}}{(\rho^2+L^2)^{\frac{7-p}{2}}}}}=0\ .\label{EOMDp/Dq}
\end{equation}
For large $\rho\to\infty$ the second term in equation~(\ref{EOMDp/Dq}) vanishes and the solution $L(\rho)$ has the asymptotic behavior:
\begin{equation} 
L(\rho)=m+\frac{c}{\rho^{n-1}}+\cdots\ ,
\end{equation}
which encodes \cite{Karch:2002sh,Kruczenski:2003uq} the bare quark
mass $m_q={m}/{2\pi\alpha'}$ and the quark bilinear condensate
$\langle\bar\psi\psi\rangle\propto-c$ of the dual gauge theory.

It is also clear that the equation of motion (\ref{EOMDp/Dq}) has a
trivial solution $L(\rho)\equiv 0$, which preserves the rotational
symmetry in the $8-p-n$ plane transverse to both the Dp and
D$q$--branes. This solution has zero bare quark mass and corresponds
to the non-symmetry breaking phase of the dual gauge theory. The
solutions in the vicinity of $L\equiv 0$ are unstable and correspond
to the interior of the spiral structure that we are studying. In order
to obtain the scaling exponents characterizing the spiral we will zoom
in on the region close to the origin of the $(\rho,L)$ plane. We first
introduce dimensionless variables {\it via}:
\begin{equation}
\rho=\tilde\rho RH^{\frac{2}{7-p}}\ ;~~L=\tilde L RH^{\frac{2}{7-p}}\ ;~~\tilde m=mRH^{\frac{2}{7-p}};~~c=\tilde c R^nH^{\frac{2n}{7-p}}\ ; 
\end{equation}
and now rescale:
\begin{equation}
\tilde \rho=\lambda \hat\rho\ ;~~~ \tilde L=\lambda\hat L\ .
\end{equation}
In the limit $\lambda\to 0$ equation (\ref{EOMDp/Dq}) becomes:
\begin{equation}
\partial_{\hat\rho}\left(\frac{{\hat\rho}^n}{(\hat\rho^2+{\hat L}^2)^{\frac{7-p}{4}}}\frac{\hat L'}{\sqrt{1+\hat L'^2}}\right)+\frac{7-p}{2}\sqrt{1+\hat L'^2}\frac{\hat\rho^n\hat L}{(\hat\rho^2+\hat L^2)^{\frac{11-p}{4}}}=0\ .\label{EOMSC}
\end{equation}
The solutions to equation (\ref{EOMSC}) have the scaling property that
if $\hat L(\hat\rho)$ is a solution, then so is $\frac{1}{\mu}\hat
L(\mu\hat\rho)$. In order to explore the vicinity of the critical
$\hat L\equiv 0$ solution we define $\hat L=0+\zeta(\hat\rho)$ and
linearize with respect to $\zeta$, the result is:
\begin{equation}
\partial_{\hat\rho}(\hat\rho^{n-\frac{7-p}{2}}\zeta')+\frac{7-p}{2}\hat\rho^{n-\frac{11-p}{2}}\zeta=0\ .\label{linsc}
\end{equation}
Next we look for solutions of equation (\ref{linsc}) of the form $\zeta=\hat\rho^{\nu}$. The quadratic equation for $\nu$ that  we obtain is:
\begin{equation}
2\nu^2+(n+d-6)\nu+(n-d+4)=0\ .\label{quadratno}
\end{equation}
We have used the constraint $p=3+d-n$. Now in order to have a
logarithmic spiral (which seeds the multi-valuedness of the equation
state) we need to have two complex roots. The condition for that is:
\begin{equation}
(n+d-6)^2<8(n-d+4)\ .\label{condition} 
\end{equation}
Note that in order to be able to turn on a magnetic field we need
$d\ge 2$. In addition we are not interested in theories with $d>3$. It
is then easy to check that for all possible values of $n$ (clearly
$n<5$) the condition (\ref{condition}) is satisfied. Then the roots of
equation~(\ref{quadratno}) $\nu_{\pm}$ are given by:
\begin{equation}
\nu_{\pm}=-r_{n,d}\pm i\alpha_{n,d}\ ; ~~~r_{n,d}=\frac{n+d-6}{4}\ge-\frac{3}{4}\ ;~~\alpha_{n,d}=\frac{1}{4}\sqrt{8(n-d+4)-(n+d-6)^2}\ .\label{exponents}
\end{equation}
The inequality in the second formula in equation~(\ref{exponents}) is
saturated for the minimum possible values $(n, d)=(1,2)$. The most
general solution of equation~(\ref{linsc}) can then be written as:
\begin{equation}
\zeta(\hat\rho)=\frac{1}{\hat\rho^{r_{n,d}}}(A\cos(\alpha_{n,d}\ln\hat\rho)+B\sin(\alpha_{n,d}\ln\hat\rho))\ .
\end{equation}
Now the scaling property of equation (\ref{EOMSC}) suggests the
following transformation of the parameters $(A,B)$ under re-scaling of
the initial condition $\hat L(0)\equiv L_0\to\frac{1}{\mu}\hat L_0$:
\begin{equation} 
\begin{pmatrix} A' \\  B' \end{pmatrix}=\frac{1}{\mu^{r_n+1}}\begin{pmatrix}\cos{(\alpha_n\ln\mu)} & \sin{(\alpha_n\ln\mu)}\\-\sin{(\alpha_n\ln\mu)} & \cos(\alpha_n\ln\mu)\end{pmatrix}\begin{pmatrix} A \\  B \end{pmatrix}\ .
\label{transformation}
\end{equation}
For a fixed choice of the parameters $A$ and $B$, the parameters $(A',
B')$ describe a logarithmic spiral, whose step and periodicity are set
by the real and imaginary parts of the critical/scaling exponents
$r_{n,d}$ and $\alpha_{n,d}$. Note that from the inequality in
equation~(\ref{exponents}) it follows that $r_{n,d}+1\ge\frac{1}{4}>0$ and
hence the spiral is revolving as one scales away from the critical
$\hat L\equiv 0$ solution.

This self--similar structure of the embeddings near the critical
solution $\hat L\equiv 0$ in our zoomed in region parameterized by
$(\hat\rho,\hat L)$ is transferred by a linear transformation to the
structure of the solutions in the $(m, c)$ parameter space.  The
parameters corresponding to the critical $L\equiv 0$ embedding are
given by $(0,0)$. Then sufficiently close to the critical embedding we
can expand:
\begin{equation}
\begin{pmatrix}
m\\c
\end{pmatrix}=M\begin{pmatrix}A\\B\end{pmatrix}+{\cal O}(A^2, B^2, AB)\ .
\end{equation}
The constant matrix $M$ depends on the properties of the
system. Generically it should be invertible (numerically we have
verified that this is the case) and therefore in the vicinity of the
parameter space close to the critical embedding $(m,c)$ there is a
discrete self--similar structure determined by the transformation:
\begin{equation}
\begin{pmatrix} m'\\  c' \end{pmatrix}=\frac{1}{\mu^{r_n+1}}M\begin{pmatrix}\cos{(\alpha_n\ln\mu)} & \sin{(\alpha_n\ln\mu)}\\-\sin{(\alpha_n\ln\mu)} & \cos(\alpha_n\ln\mu)\end{pmatrix}M^{-1}\begin{pmatrix} m\\  c\end{pmatrix}\ . 
\label{fultrans}
\end{equation}
 
Note that the linear map corresponding to the constant matrix $M$
would rotate, stretch and/or shrink (along the different axes) the
spiral defined {\it via} the transformation
(\ref{transformation}). However the overall shape of the curve defined
{\it via} equation~(\ref{fultrans}) still remains a spiral revolving
around the origin of the $(\tilde m, \tilde c)$ plane (see
figure~\ref{fig:fig3} for the case of the D3/D7 system). This suggests
that the state corresponding to the center of the spiral (the $L\equiv
0$ solution) is unstable and hence there is a dynamical mass
generation in the theory. (The stable state at zero bare quark mass
has a non--zero condensate) Therefore we learn that for all Dp/Dq
systems T--dual to the D3/D7 intersection (and with $d\ge 2$ so that a
magnetic field can be switched on) the effect of the magnetic field is
to break a global internal symmetry and generate a dynamical mass.
 
\begin{figure}[h] 
   \centering
   \includegraphics[width=10cm]{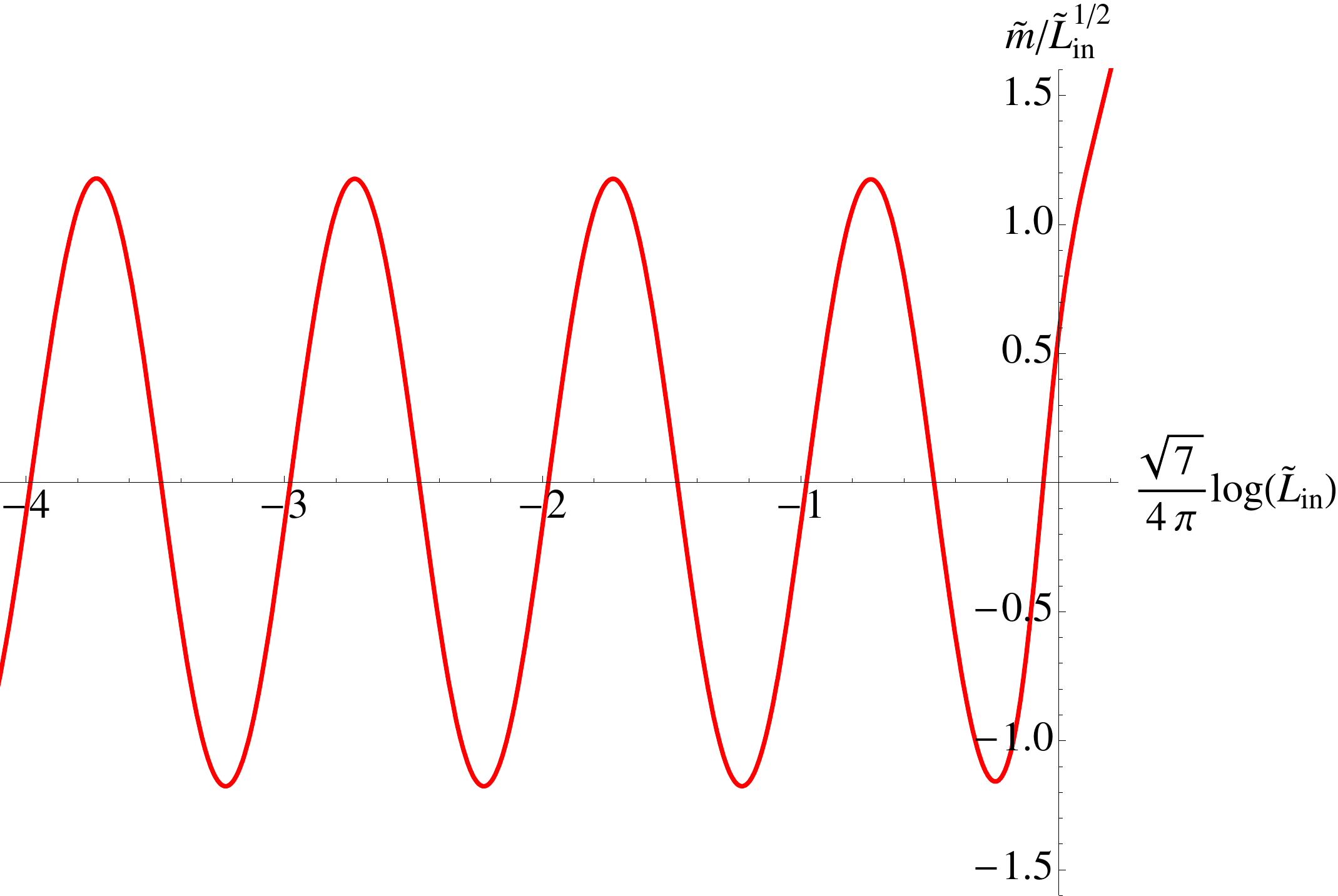}
   \caption{\small A plot of $\frac{\sqrt{7}}{4\pi}\log\tilde L_{in}$ {\it vs.} $\tilde m/\tilde L_{in}^{1/2}$. For sufficiently small $\tilde L_{in}$ the curve is an harmonic function of unit period.}
\label{fig:fig9}
\end{figure}

To conclude this discussion we will provide a numerical check of the
consistency of our analysis. To this end we consider the separation of
the Dq and Dp branes at $L_{in}\equiv L(0)$ (note that $L_{in}$ is
proportional to the dynamically generated quark mass). Now if we start
from some $L^0_{in}$ and transform to $L_{in}=\frac{1}{\mu}L^0_{in}$,
we can solve for $\mu$ and generate a parametric plot of $\tilde
m/(\tilde L_{in})^{r_{n,d}+1}$ {\it vs.} $\alpha_{n,d}\log\tilde
L_{in}/2\pi$. The transformation~(\ref{fultrans}) requires that the
resulting plot be an harmonic function of unit period. For the
particular case of the D3/D5 system we have
$r_{2,2}=-1/2$ and $\alpha_{2,2}=\sqrt{7}/2$. The corresponding plot
is presented in figure~\ref{fig:fig9}.  For sufficiently small $\tilde
L_{in}$ the plot is indeed an harmonic function of unit period.
 
\section{Conclusion}
 
 In this paper we have investigated the properties of strongly coupled, large $N_c$ gauge theories in the presence of an external magnetic field. Both in three and four dimensions, the holographic approach reproduces the behavior expected from classical field theory arguments and the magnetic catalysis of global symmetry breaking is shown to be a universal feature of a family of strongly coupled gauge theories. This is another success for the internal consistency of the gauge/gravity duality.

As has now become commonplace in such studies, the spectrum of mesonic states is directly identifiable by numerical methods. Moreover, in the present study we are able to make a large number of statements analytically by studying the systems in the chiral limit. Indeed in this regime, in which the global symmetry is broken dynamically due to the presence of the magnetic field the results are almost completely tractable analytically. 

In $3+1$ dimensions we are able to show that the Gell-Mann--Oakes--Renner  relation holds exactly and in this setting where a subgroup of the special relativistic transformations remains unbroken in the presence of the magnetic field, a relativistic dispersion relation is indeed recovered. In the $2+1$ dimensional case however, all trace of the boost invariance is lost once the magnetic field is turned on. The counting of Goldstone modes then becomes more subtle but we are able to show that this holographic setup gives the correct number of massless modes expected from the non-relativistic Goldstone counting rules. In addition we are able to show that these modes obey a quadratic dispersion relation, in contrast to the relativistic case in one spatial dimension higher.

In the present investigation we have focused on the low energy chiral lagrangian, calculated up to quadratic order in the goldstone mode excitations. In the $2+1$ dimensional setting we were able to show explicitly that at this order our system reproduces the low energy behavior of spin-wave excitations in a ferromagnet. The matching of symmetry breaking patterns between the two systems is the root of this equality. It seems unlikely however that such an agreement will hold to higher order. It would certainly be possible to investigate the systems discussed here at higher order in the low energy degrees of freedom and this would be an interesting direction for future work.

The study of flavor degrees of freedom in a wide range of conditions is certainly well worth pursuing further using the methods illustrated in this paper. The magnitude of the phase-space involved with studying flavor, both abelian and non-abelian, in the presence of external electric fields and magnetic fields, temperature and finite chemical potential in a variety of dimensions means that there is surely plenty more to be found in even the simplest systems.

In the case of $2+1$ dimensions, the understanding of gauge theories in the presence of low temperature and high magnetic field is of much interest. In particular this is one area which may be accessible experimentally and one may be able to gain insight into such intriguing phenomena as the quantum hall effect. It seems very likely that we will make further inroads into understanding such effects using the AdS/CFT correspondence in the not too distant future. 

The diversity of phenomena that we can investigate using holographic techniques is clearly far larger than was expected in the early days of the AdS/CFT conjecture. The prospects of obtaining deep insight into such fascinating systems as non-conventional superconductors, the quantum hall effect and strongly coupled plasmas are real and exciting and the community continues to make progress in these directions.

\section{Acknowledgements}
We would like to thank B. Doulen, N. Evans, D. O'Connor, J. Erdmenger, P. Kraus, B. Queshi, R. Meyer, C. Nunez and A. O'Bannon  for useful comments and suggestions. V.F. would like to thank University of Santiago de Copmostela and Max Planck Institute for Physics for their hospitality at the final stage of this project. The work of V.F. was
supported by a IRCSET fellowship. The work of C.V.J. was supported by the
US Department of energy. The work of J.S was supported in part by
MICINN and FEDER (grant FPA2008-01838), by Xunta de Galicia
(Conseller'a de Educaci—n and grant PGIDIT06PXIB206185PR), and
by the Spanish Consolider-Ingenio 2010 Programme CPAN
(CSD2007-00042). J.S is supported by a Juan de la Cierva fellowship.

\bigskip
\bigskip
\section*{Appendices}
\appendix \section{Low energy effective action for $\Phi$} \label{appendix:A}

Let us begin with the Lagrangian for the quadratic fluctuations equation~(\ref{qvd2})
\begin{eqnarray}
&&{\cal L}_{\Phi\Phi}=-(2\pi\alpha')^2\frac{\mu_7}{g_s}\frac{1}{2}\sqrt{|g_{S^3}|}\frac{gR^2L_0^2}{\rho^2+L_0^2}S^{ab}\partial_a\Phi\partial_b\Phi\, , \label{qvd2app}\\
&&{\cal L}_{\Phi A}=-(2\pi\alpha')^2\frac{\mu_7}{g_s}\sqrt{|g_{S^3}|}H\partial_{\rho}K\Phi F_{01}\ ,\nonumber\\
&&{\cal L}_{AA}=-(2\pi\alpha')^2\frac{\mu_7}{g_s}\sqrt{|g_{S^3}|}\frac{1}{4}gS^{aa'}S^{bb'}F_{ab}F_{a'b'}\ .\nonumber
\end{eqnarray}
In order to obtain an effective action for the fluctuations along
$\Phi$ we will integrate out the fluctuations of the gauge field and
in particular the $A_0,A_1$ components. In more detail the
contribution from the last two equations in equation~(\ref{qvd2app}) can be
written as:
\begin{eqnarray}
{\cal L}_{AA}+{\cal L}_{\Phi A}&\propto&\frac{1}{2}gS^{11}S^{aa'}(-\partial_aA_0\partial_{a'}A_0+\partial_aA_1\partial_{a'}A_1)-\frac{1}{2}g(S^{11})^2(\partial_0A_0-\partial_1A_1)^2\label{flA2}\\
&&\hskip1cm+H\partial_{\rho}K\Phi F_{01}\ .\nonumber
\end{eqnarray}
To integrate out the $A_0,A_1$ components of the gauge field we simply
obtain the equations of motion for $A_0$ and $A_1$ and substitute them
into the action. The equations of motion are:
\begin{eqnarray}
&&\partial_a(gS^{11}S^{aa'}\partial_aA_0)+g(S^{11})^2\partial_0(\partial_0A_0-\partial_1A_1)+H\partial_{\rho}K\partial_1\Phi=0\ ,\\
&&\partial_a(gS^{11}S^{aa'}\partial_aA_1)+g(S^{11})^2\partial_1(\partial_0A_0-\partial_1A_1)+H\partial_{\rho}K\partial_0\Phi=0\ ,\nonumber
\end{eqnarray}
which can be written as:
\begin{eqnarray}
&&\partial_a\left(gS^{11}S^{aa'}\partial_aF_{01}\right)-H\partial_{\rho}K\Box_{(1,1)}\Phi=0\ ;~~~\Box_{(1,1)}\equiv-\partial_0^2+\partial_1^2\ ,\\
&&\partial_a\left(gS^{11}S^{aa'}\partial_a(\partial_0A_0-\partial_1A_1)\right)-g(S^{11})^2\Box_{(1,1)}(\partial_0A_0-\partial_1A_1)=0\nonumber\ .
\end{eqnarray}
Substituting back into the action (\ref{flA2}) and integrating by
parts leads to:
\begin{equation}
{\cal L}_{AA}+{\cal L}_{\Phi A}\propto-\frac{1}{2}HK\partial_{\rho}(\Phi F_{01})\, .\label{actINT}
\end{equation}
The equation of motion for $F_{01}$ can be written as:
\begin{equation}
\partial_\rho F_{01}=\frac{HK}{\Psi_1^2}\Box_{(1,1)}\Phi-\frac{1}{\Psi_1^2}\int d\rho g(S^{11})^2\tilde\Box F_{01}\ ,
\end{equation}
where $\Psi_1$ is defined in equation (\ref{PSIS}). Now we substitute
into the action (\ref{actINT}) and denote by $\tilde\Phi(x)$ the
dimensionally reduced field $\Phi$, to obtain:
\begin{equation}
{\cal L}_{AA}+{\cal L}_{\Phi A}=(2\pi\alpha')^2\frac{\mu_7}{g_s}\sqrt{|g_{S^3}|}\frac{1}{2}\frac{H^2K^2}{\Psi_1^2}\tilde\Phi\Box_{(1,1)}\tilde\Phi+\dots\ ,\label{ACT3}
\end{equation}
where we have ignored higher order derivatives terms. Combining this
with the dimensionally reduced term ${\cal L}_{\Phi\Phi}$ we obtain
the result:
\begin{eqnarray} {\cal
    L}\propto\frac{1}{2}(\nu\Psi^2+\frac{H^2K^2}{\Psi_1^2})[-(\partial_0\tilde\Phi)^2+(\partial_1\tilde\Phi)^2]+\frac{1}{2}\tilde\nu\Psi^2[(\partial_2\tilde\Phi)^2+(\partial_3\tilde\Phi)^2]+\dots\, .\label{almostfin}
\end{eqnarray}
Where $\nu,\tilde\nu$ and $\Psi$ are defined in equation
(\ref{PSIS}). In order to obtain a mass term for the dimensionally
reduced field $\tilde\Phi$ we have to take into account the radial
dependence of the field $\Phi$. Our analysis from Section 2.2.1
suggests that we should consider the following ansatz:
\begin{equation}
\Phi(\rho,x)=\frac{\psi(\rho)}{\Psi(\rho)}\tilde\Phi(x)\ ,
\end{equation}
where $\Psi$ is defined in equation~(\ref{PSIS}) and we require that
for the spontaneous symmetry breaking classical embedding (denoted by
$\bar L_0$) we have that $\psi|_{\bar L_0}=\Psi|_{\bar L_0}\equiv
\bar\Psi$. Then if we consider embeddings in the vicinity of $\bar
L_0$ corresponding to small bare quark mass $\delta m$ we can expand:
\begin{equation}
\psi=\bar\Psi+\delta\psi\ ;~~~\Phi(\rho,x)=\left[1+\delta\left(\frac{\psi}{\Psi}\right)\right]\tilde\Phi(x)\ .
\end{equation}
Now if we demand that as $\delta m\to0$ we have small momenta and a
small mass term (which vanish at the critical embedding) to leading
order we still have the expression from equation (\ref{almostfin})
plus some small mass term involving derivatives along $\rho$:
\begin{eqnarray}
{\cal L}&\propto& \frac{1}{2}(\bar\nu{\bar\Psi}^2+\frac{H^2\bar K^2}{{\bar\Psi_1}^2})[-(\partial_0\tilde\Phi)^2+(\partial_1\tilde\Phi)^2]+\frac{1}{2}\bar{\tilde\nu}{\bar\Psi}^2[(\partial_2\tilde\Phi)^2+(\partial_3\tilde\Phi)^2]\nonumber\\
&&\hskip1cm -\frac{1}{2}\partial_{\rho}\left[\bar\Psi^2\partial_{\rho}\delta\left(\frac{\psi}{\Psi}\right)\right]\tilde\Phi^2+\dots\ ,
\end{eqnarray}
where we have integrated by parts the last term and the dots represent
higher derivatives terms and other sub-leading terms. Now it is straightforward to integrate along the unit $S^3$ and the
radial coordinate $\rho$. Let us provide some more details in the
integration of the mass term:
\begin{eqnarray}
\int\limits_{0}^{\infty}d\rho\partial_{\rho}\left[\bar\Psi^2\partial_{\rho}\delta\left(\frac{\psi}{\Psi}\right)\right]\tilde\Phi^2=\left[(\bar\Psi\delta\psi'-\delta\psi\bar\Psi')+(\delta\Psi{\bar\Psi}'-\delta\Psi' \bar\Psi)\right]\tilde\Phi^2\Big|_0^{\infty}=-2c\delta m\tilde\Phi^2\, .
\end{eqnarray}
Then for the final form of the effective action one obtains:
\begin{equation}
S_{\rm{eff}}=-{\cal N}\int d^4x\left\{[-(\partial_0\tilde\Phi)^2+(\partial_1\tilde\Phi)^2]+\gamma[(\partial_2\tilde\Phi)^2+(\partial_3\tilde\Phi)^2]-\frac{2\langle\bar\psi\psi\rangle}{f_{\pi}^2}m_q\tilde\Phi^2\right\}+\dots\ ,\label{appeneff}
\end{equation}
where:
\begin{eqnarray}
&&{\cal N}=(2\pi\alpha')^2N_f\frac{\mu_7}{g_s}\pi^2\int\limits_0^{\infty}d\rho\left[\bar\nu{\bar\Psi}^2+\frac{H^2\bar K^2}{{\bar{\Psi}_1}^2}\right];~~~f_{\pi}^2=\frac{4{\cal N}}{(2\pi\alpha')^2};~~~m_q=\frac{\delta m}{2\pi\alpha'}\, ,\label{appendconst}\\
&&\gamma=\int\limits_0^{\infty}d\rho\left(\bar{\tilde\nu}{\bar\Psi}^2\right)\Big/\int\limits_0^{\infty}d\rho\left(\bar\nu{\bar\Psi}^2+\frac{H^2\bar K^2}{\bar{\Psi_1}^2}\right);~~~\langle\bar\psi\psi\rangle=-\frac{N_f}{(2\pi\alpha')^3}\frac{c}{2\pi g_s}\, .
\end{eqnarray}

One can see that this is the most general quadratic action consistent with the $SO(1,1)\times SO(2)$ space-time symmetry. Furthermore the explicit form of the mass term is in accord with the Gell-Mann--Oakes--Renner relation (\ref{GellMann2}). To obtain the expression for $f_{\pi}^2$ provided in equation (\ref{appendconst}) one needs to consider the strict $m_q\to0$ limit and use that in this limit $\tilde\Phi=\phi/(2\pi\alpha')$. Next since $\phi$ corresponds to rotations in the transverse $\IR^2$ plane and is thus the angle of chiral rotation~\cite{Kruczenski:2003uq}, the normalization of the kinetic term in the effective action (\ref{appeneff}) is given by ${\cal N}=(2\pi\alpha')^2f_{\pi}^2/4$. The last relation determines $f_{\pi}^2$ in terms of ${\cal N}$.

 
\end{document}